\documentclass[10pt,journal]{IEEEtran}
\IEEEoverridecommandlockouts
\usepackage{amssymb}
\usepackage{amsfonts}
\usepackage{epstopdf}
\usepackage{graphicx}
\usepackage{stmaryrd}
\usepackage{amsmath}
\usepackage{multirow,url}
\usepackage{color,cite}
\usepackage{hyperref} 
\usepackage{verbatim} 
\usepackage[OT1]{fontenc} 
\usepackage{titlesec}
\usepackage{algorithm}
\usepackage{algorithmic}
\usepackage{subcaption}
\usepackage{textcomp}

\usepackage{caption}  
\begin{document}

\title{Diffusion Model-based Reinforcement Learning for \\ Version Age of Information Scheduling: \\Average and Tail-Risk-Sensitive Control}

\author{Haoyuan~Pan,~\IEEEmembership{Member,~IEEE,}~Sizhao~Chen,~Zhaorui~Wang,~\IEEEmembership{Member,~IEEE,}~Tse-Tin Chan, \IEEEmembership{Member, IEEE}%
\thanks{H. Pan and S. Chen are with the College of Computer Science and Software Engineering, Shenzhen University, Shenzhen, China (e-mails:  {hypan@szu.edu.cn}, {2510103060@mails.szu.edu.cn}).}
\thanks{Z. Wang is with the School of Computing and Information Technology, Great Bay University, Dongguan, China (e-mail: wangzhaorui@gbu.edu.cn).}
\thanks{T.-T. Chan is with the Department of Mathematics and Information Technology, The Education University of Hong Kong, Hong Kong, China (e-mail: tsetinchan@eduhk.hk).}
}

\maketitle

\begin{abstract}
Ensuring timely and semantically accurate information delivery is critical in real-time wireless systems. While Age of Information (AoI) quantifies temporal freshness, Version Age of Information (VAoI) captures semantic staleness by accounting for version evolution between transmitters and receivers. Existing VAoI scheduling approaches primarily focus on minimizing average VAoI, overlooking rare but severe staleness events that can compromise reliability under stochastic packet arrivals and unreliable channels. This paper investigates both average-oriented and tail-risk-sensitive VAoI scheduling in a multi-user status update system with long-term transmission cost constraints. We first formulate the average VAoI minimization problem as a constrained Markov decision process and introduce a deep diffusion-based Soft Actor-Critic (D2SAC) algorithm. By generating actions through a diffusion-based denoising process, D2SAC enhances policy expressiveness and establishes a strong baseline for mean performance. Building on this foundation, we put forth RS-D3SAC, a risk-sensitive deep distributional diffusion-based Soft Actor-Critic algorithm. RS-D3SAC integrates a diffusion-based actor with a quantile-based distributional critic, explicitly modeling the full VAoI return distribution. This enables principled tail-risk optimization via Conditional Value-at-Risk (CVaR) while satisfying long-term transmission cost constraints. Extensive simulations show that, while D2SAC reduces average VAoI, RS-D3SAC consistently achieves substantial reductions in CVaR without sacrificing mean performance. The dominant gain in tail-risk reduction stems from the distributional critic, with the diffusion-based actor providing complementary refinement to stabilize and enrich policy decisions. These results highlight the effectiveness of combining diffusion-based policy learning with distributional reinforcement learning for robust and risk-aware VAoI scheduling in multi-user wireless systems.
\end{abstract}


\IEEEpeerreviewmaketitle

\section{Introduction}
In real-time wireless communication systems, maintaining the timeliness of information updates is crucial for system perception, control, and decision-making. To quantitatively characterize information freshness, the Age of Information (AoI) has been widely adopted as a metric that measures the time elapsed since the generation of the most recently received update \cite{AoI_apply_1}. AoI effectively captures the temporal freshness of information and has therefore been extensively studied in a wide range of communication scenarios, including status update systems, networked control, and cyber-physical systems \cite{AoI_apply}, \cite{AoI_apply_2}. 

With the rapid emergence of semantic and task-oriented communication paradigms, however, characterizing information freshness solely from a temporal perspective is no longer sufficient. In many practical systems, information evolves not only over time but also across different versions, where outdated semantic content may severely degrade system performance even when updates are timely. To capture this semantic evolution, the Version Age of Information (VAoI) metric has been introduced to quantify the version discrepancy between transmitters and receivers \cite{VAoI}, \cite{VAoI_1}, \cite{VAoI_2}. Compared with conventional AoI, VAoI provides a more expressive and semantically meaningful measure of information freshness, offering a richer foundation for scheduling and control design in wireless communication networks.

Existing work on VAoI optimization and scheduling has predominantly adopted a \emph{risk-neutral} formulation that minimizes the long-term average VAoI to improve overall semantic freshness \cite{average_vaoi, average_vaoi_1}. A similar average-oriented methodology is also widely used in the traditional AoI literature \cite{AoI_apply_2}. While effective in improving mean performance, such formulations do not explicitly characterize or control the variability and tail behavior of information staleness. Unlike classical AoI, where large values indicate temporal staleness, high VAoI indicates prolonged semantic invalidity of information. In stochastic wireless systems with random packet arrivals and unreliable channels, rare but severe VAoI events can thus lead to incorrect semantic inference or unreliable downstream decisions, rendering the system vulnerable to catastrophic semantic failures even when the average VAoI is low.

These observations motivate a \emph{risk-sensitive} VAoI framework that explicitly controls tail behavior, while retaining average-performance considerations, thereby complementing existing mean-oriented approaches. In particular, the Conditional Value-at-Risk (CVaR) emerges as a principled tail-risk metric that characterizes the expected performance under the worst fraction of realizations \cite{CVaR}. By explicitly accounting for extreme outcomes, CVaR provides a more robust and interpretable measure of semantic staleness risk, making it particularly suitable for mission-critical and safety-sensitive applications \cite{CVaR_apply_1}. Importantly, CVaR characterizes the tail behavior of the underlying performance metric, implying that optimization approaches based solely on expectation are inherently insufficient for controlling rare but severe events.

From a system perspective, VAoI scheduling problems are further complicated by stochastic data arrivals, time-varying wireless channels, and long-term resource limitations. These factors give rise to complex system dynamics that are difficult to characterize accurately using analytical models. Classical model-based optimization approaches \cite{model_based_method}, \cite{model_based_method_1} typically require complete system knowledge and often struggle to achieve scalability and robustness in such uncertain environments. In contrast, deep reinforcement learning (DRL) has recently demonstrated strong potential for wireless resource allocation and information freshness optimization, thanks to its powerful representation capabilities and data-driven nature \cite{DRL_apply}. In particular, model-free DRL methods learn control policies directly from interaction data without explicit system models, making them well-suited for VAoI scheduling problems with unknown or time-varying dynamics.

Despite their empirical success, most existing DRL-based VAoI scheduling approaches are inherently expectation-based and thus offer limited capability to control tail risks associated with rare but severe semantic staleness events. Moreover, conventional policy networks often lack sufficient expressive power to accurately represent the complex and highly structured action distributions induced by discrete scheduling decisions and stringent wireless resource constraints. These limitations motivate the exploration of more expressive policy representations and risk-aware learning paradigms. Recently, diffusion models have demonstrated remarkable effectiveness in modeling complex distributions and have been successfully integrated into DRL frameworks to enhance policy expressiveness and training stability \cite{diffusion_drl}, \cite{diffusion_drl_1}. By generating actions through a progressive denoising process, diffusion-based policies can capture rich structural dependencies in the action space, making them particularly well-suited for challenging wireless scheduling problems. 

In parallel, distributional reinforcement learning departs from classical expectation-based value estimation by explicitly modeling the entire return distribution, thereby providing a principled foundation for tail-risk-aware optimization via coherent risk measures such as CVaR \cite{qrdqn}. Prior studies have shown that distributional DRL can significantly outperform conventional DRL methods in wireless resource allocation and radio management scenarios, particularly when variability and uncertainty are  dominant \cite{dis_DRL_1}. These observations suggest that combining expressive diffusion-based policies with distribution-aware value estimation is a key enabler for effective risk-sensitive VAoI scheduling.

Motivated by these observations, we consider a multi-user wireless status update system with stochastic packet arrivals and random transmission success probabilities. Under the risk-neutral framework, we formulate the problem of minimizing the long-term average VAoI subject to long-term transmission cost constraints as a Constrained Markov Decision Process (CMDP) \cite{CMDP}. To address this problem, we develop a constrained deep diffusion-based Soft Actor-Critic (D2SAC) algorithm \cite{diffusion_d2sac}, where the diffusion process enhances the policy representation capability and enables efficient optimization of the average VAoI.

Building upon this foundation, we further investigate risk-sensitive VAoI scheduling by incorporating CVaR as the tail-risk metric. Leveraging quantile-based distributional modeling techniques inspired by QR-DQN  \cite{qrdqn}, we put forth a risk-sensitive deep distributional diffusion-based Soft Actor-Critic (RS-D3SAC) algorithm. By explicitly learning the distribution of VAoI returns and coupling it with diffusion-based policy generation, RS-D3SAC enables principled control of extreme semantic staleness events without violating long-term transmission cost constraints.

Extensive simulation results demonstrate that the proposed algorithms significantly outperform existing baselines in terms of both average VAoI and tail-risk mitigation. In particular, D2SAC achieves substantial improvements in average VAoI compared with standard SAC and other conventional DRL approaches, validating the effectiveness of diffusion-enhanced policy representations. More importantly, owing to the joint integration of a diffusion-based actor and a distributional critic, RS-D3SAC attains significantly lower CVaR of VAoI than D2SAC and all benchmark algorithms. Notably, the dominant gains in tail-risk reduction stem from the distributional critic, while the diffusion-based actor provides complementary refinement through improved action expressiveness.

The main contributions of this paper are summarized as follows:
\begin{itemize}
    \item We formulate VAoI scheduling in a multi-user wireless status update system with stochastic arrivals and unreliable channels as a CMDP. Under the risk-neutral framework, we develop a constrained D2SAC algorithm to efficiently minimize the long-term average VAoI subject to transmission cost constraints.
    
    \item We move beyond expectation-oriented optimization and propose a risk-sensitive VAoI scheduling framework based on CVaR. To this end, we put forth the RS-D3SAC algorithm, which integrates a quantile-based distributional critic with a diffusion-based actor to enable principled tail-risk control.
    
    \item Through extensive simulations, we demonstrate that D2SAC substantially improves average VAoI compared with standard SAC and other DRL baselines, while RS-D3SAC achieves significant reductions in VAoI tail risk. The results further reveal that the dominant gain in tail-risk mitigation originates from the distributional critic, with the diffusion-based actor providing complementary performance refinement.
\end{itemize}


\section{Related Work}\label{sec:related_work}
\subsection{Semantic Freshness Metrics and Scheduling Approaches}
AoI has become a fundamental metric for characterizing information timeliness in wireless status update systems, measuring the elapsed time since the most recently received update was generated \cite{AoI_apply_1}. It has been extensively studied in scenarios where timely information delivery is critical. We refer readers to the monograph \cite{AoI_apply_2} and the references therein for comprehensive overviews of AoI theory and applications. 

To move beyond purely temporal freshness, several semantic- and task-aware metrics have been proposed. In particular, VAoI extends AoI by explicitly accounting for version mismatches between transmitters and receivers, thereby capturing semantic staleness in systems where information evolves across discrete versions \cite{VAoI}, \cite{VAoI_1}, \cite{VAoI_2}. Related metrics, such as Urgency of Information (UoI) \cite{UoI} and Goal-Oriented Tensor (GoT) \cite{GoT}, further emphasize task-dependent or goal-oriented freshness requirements.

Under these metrics, a large body of work has focused on scheduling policies that minimize long-term average freshness. Model-based approaches have been developed for VAoI optimization in a variety of settings, including energy-harvesting gossip networks \cite{VAoI}, \cite{VAoI_1}, federated learning systems \cite{average_vaoi}, and fading broadcast channels \cite{model_based_method_1}. More recently, \cite{cvar_aoi} incorporated CVaR into AoI optimization to mitigate rare but severe staleness events, while \cite{StatisticalAoI} proposed statistical AoI as a tunable metric interpolating between average and peak age. Although these methods provide valuable analytical insights and performance guarantees, they typically rely on precise system models and simplifying assumptions, which limit scalability and robustness in highly dynamic wireless environments. Moreover, to the best of our knowledge, no systematic risk-sensitive study of VAoI has explicitly targeted the semantic staleness tail behavior.

To address modeling limitations, reinforcement learning has been introduced as a model-free alternative for VAoI scheduling. For example, Q-learning-based methods have been applied to energy-harvesting IoT systems to learn adaptive scheduling policies under uncertainty \cite{vaoi_rl}. However, existing learning-based approaches remain largely risk-neutral, focusing on minimizing average VAoI. While a few studies \cite{risk_drl_aoi_1} \cite{risk_drl_aoi_2} have applied DRL to risk-sensitive AoI objectives, the fundamental distinction between AoI and VAoI prevents these methods from explicitly controlling rare but severe semantic staleness events. Such events can disproportionately affect reliability and safety in stochastic wireless systems, motivating the adoption of tail-risk-aware metrics such as CVaR for VAoI scheduling, as pursued in this work.

\subsection{Diffusion Model-Based Reinforcement Learning}
Diffusion models have recently achieved remarkable success in generative modeling and have been introduced into reinforcement learning to enhance policy expressiveness and exploration capability \cite{zhu2023diffusion}, \cite{diffusion_DQL}. In continuous action spaces, Diffusion Q-Learning (DQL) employs conditional diffusion models to represent policies and enables efficient policy improvement in offline reinforcement learning settings \cite{diffusion_DQL}. Beyond continuous control, diffusion-based policy generation has been extended to discrete and structured decision spaces relevant to communication networks. The work in \cite{diffusion_d2sac} integrated diffusion models into the Soft Actor-Critic (SAC) framework to generate optimal decisions for AI-generated content service providers. Diffusion-based policies have also been applied to cloud-edge collaborative scheduling driven by large language model (LLM) requests \cite{diffusion_llm}, as well as satellite edge computing, where diffusion-enhanced DRL improves decision quality and generalization \cite{diffusion_computation}.

Although these studies demonstrate the strong capability of diffusion models to capture complex and constrained action distributions, existing diffusion-based DRL methods primarily focus on average performance optimization and do not explicitly consider risk-sensitive objectives. In particular, their application to information freshness optimization, especially to VAoI-aware scheduling under tail-risk constraints, remains largely unexplored.

\subsection{Distributional Reinforcement Learning and Risk-Sensitive Wireless Optimization}
Distributional DRL departs from classical RL by modeling the entire return distribution rather than only its expectation, thereby providing a richer characterization of uncertainty in stochastic environments. Representative algorithms include C51 \cite{c51}, which discretizes the return distribution support, and QR-DQN \cite{qrdqn}, which parameterizes the distribution using quantile functions and offers improved numerical stability.

By explicitly capturing tail behavior, distributional DRL has shown clear advantages in risk-sensitive decision-making problems \cite{dis_risk}. CVaR has emerged as a principled tail-risk metric and has been incorporated into DRL frameworks to improve robustness under adverse conditions \cite{lim2022distributional}. Representative works include CVaR-optimized policy gradient methods \cite{cvar_apply_3} and constrained policy optimization with CVaR-aware objectives \cite{CVaR_apply_1}. In wireless systems, these techniques have been applied to scenarios involving channel or interference uncertainty, and latency variability \cite{dis_DRL_1}, \cite{Constrained_RL}, \cite{distributionalRL_delay}.

Despite these advances, the systematic integration of distributional reinforcement learning with VAoI-aware wireless scheduling remains underexplored. In particular, existing works do not jointly leverage expressive policy representations of diffusion models and distributional value modeling to achieve scalable and principled tail-risk control for semantic information freshness. By explicitly modeling and controlling the tail behavior of VAoI under long-term transmission cost constraints, the proposed RS-D3SAC provides a principled link between semantic-aware freshness optimization and risk-sensitive reinforcement learning.

\section{System Model and Performance Metrics} \label{sec: System Model and Performance Metrics}
This section presents the system model and the Version Age of Information (VAoI) metric used to quantify information staleness in multi-user status update systems. Building on these foundations, we formalize two system-level performance criteria: (i) a risk-neutral metric, the long-term average VAoI, which captures steady-state freshness; and (ii) a risk-sensitive metric, the Conditional Value-at-Risk (CVaR) of VAoI, which accounts for extreme staleness events.

\subsection{System Model}
\begin{figure}[t]
    \centering
    \includegraphics[width=0.8\linewidth]{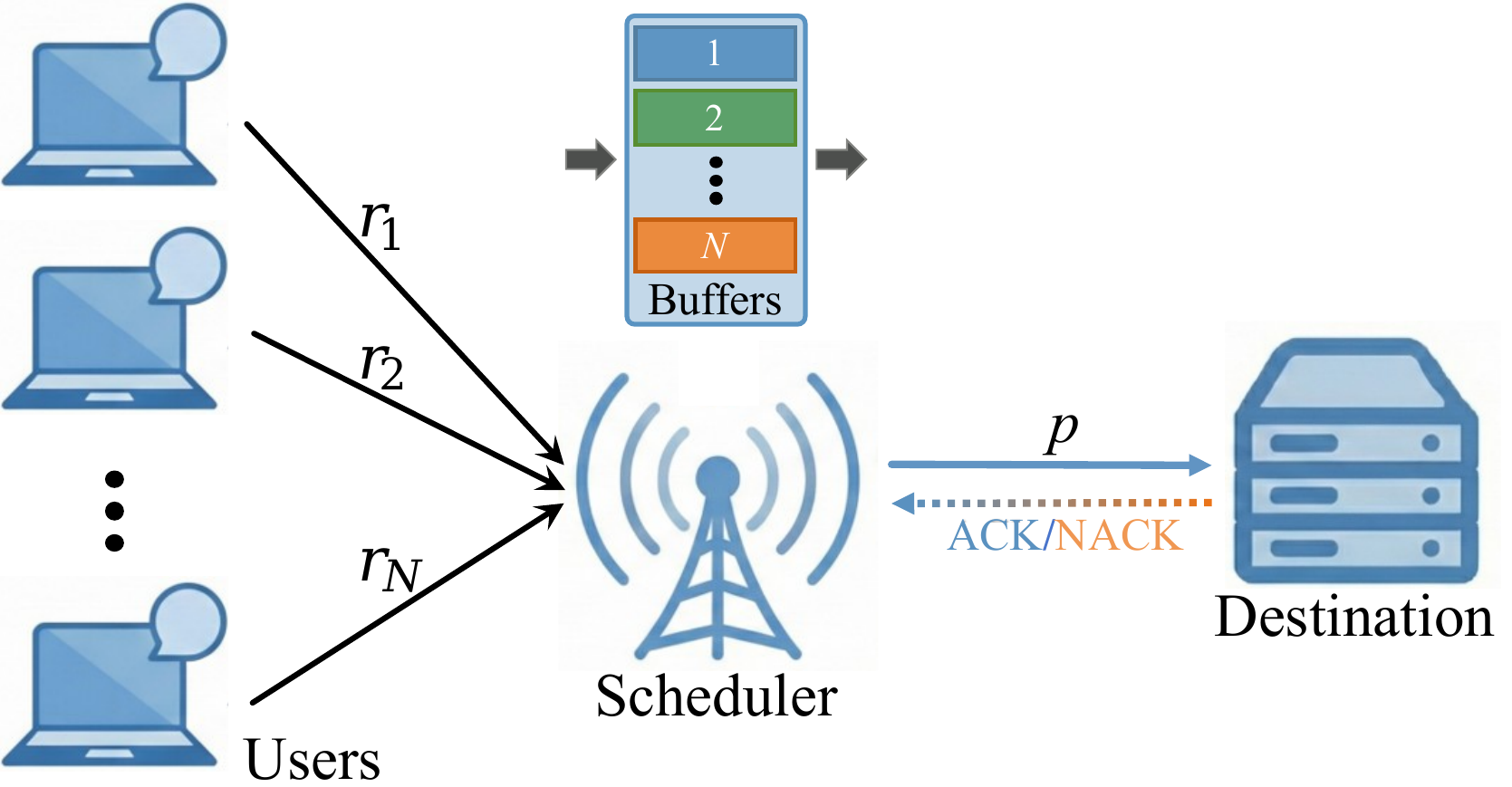}
    \caption{System model: the scheduler receives packets from 
$N$ independent users and forwards them to the destination over unreliable wireless channels, selecting at most one user per time slot.}
    \label{fig:system_model}
\vspace{-0.1in}
\end{figure}

We consider a discrete-time multi-user status update system, as illustrated in Fig. \ref{fig:system_model}. A central scheduler receives state updates from $N$ independent users and forwards them to a destination over a shared wireless channel. Time is slotted, and all slots have equal duration. At the beginning of slot $t$, user $n\in\{1,\ldots,N\}$ generates a new status packet with probability $r_n\in(0,1]$. Packet generation is instantaneous, and newly generated packets arrive at the scheduler without delay or error. The scheduler maintains a dedicated buffer for each user, storing only the most recently generated packet (i.e., $N$ buffers in total). Whenever a new packet arrives, the previous packet for that user is overwritten to prevent the accumulation of stale information.

The wireless channel from the scheduler to the destination is assumed to support at most one transmission per slot. We assume the packet duration occupies the entire slot. The scheduler must determine which user, if any, to transmit to. We define the scheduling action at slot $t$ as $a_t \in \mathcal{A} = \{0,1,\ldots,N\}$, where $a_t = 0$ represents idling and $a_t = n$ indicates that the latest packet of user $n$ is selected for transmission. A scheduled transmission succeeds with probability $p \in (0,1]$. At the end of each slot, a reliable acknowledgment (ACK) or negative ACK (NACK) is returned to the scheduler, enabling its real-time version tracking at the destination. This model captures key stochastic elements, i.e., random packet arrivals and unreliable wireless transmissions, that jointly determine information staleness in real-time systems.

\subsection{Performance Metrics}
To quantify the freshness of information delivered to the destination, we adopt the VAoI as our primary performance metric. Unlike classical AoI, which measures the elapsed time since the generation of the last successfully received update, VAoI directly tracks semantic version gaps between the scheduler and the destination. The VAoI metric therefore reflects the number of missing updates rather than the elapsed time since the last one. Let $G(t,n)$ denote the version index of user $n$'s newest packet stored at the scheduler at the beginning of slot $t$. If user $ n$ generates a new packet at slot $t$, the version stored at the scheduler increments as $G(t,n)=G(t-1,n)+1$, and remains unchanged otherwise, i.e., $G(t,n)=G(t-1,n)$. Let $B(t,n)$ denote the version index of the most recently delivered packet for user $n$ at the destination at the end of slot $t$. If user $n$ is scheduled at slot $t$ and the transmission succeeds, the version index at the destination is updated to match the one at the scheduler, $B(t,n)=G(t,n)$; otherwise, $B(t,n)=B(t-1,n)$. 

Accordingly, the instantaneous VAoI of user $n$ at slot $t$ is defined as
\begin{align}\label{formula:VAoI_Definition}
     \mathcal{D}(t,n)=G(t,n)-B(t-1,n).
 \end{align}
which captures how many update versions the destination is behind the scheduler. VAoI increases when new packets are generated or when scheduled transmissions fail, and decreases only upon successful delivery of the most recent version.

 \begin{figure}[t]
     \centering
     \includegraphics[width=1\linewidth]{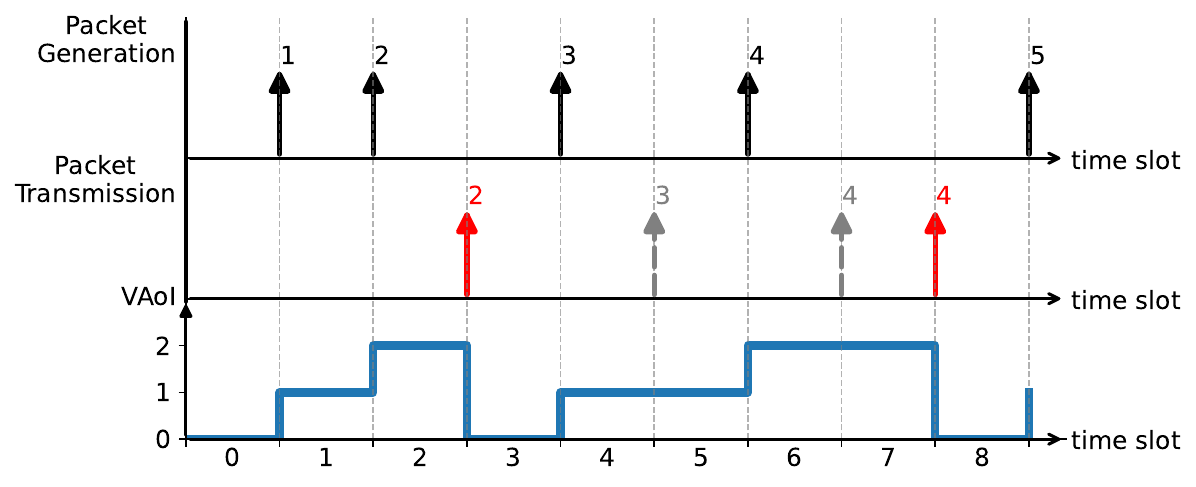}
     \caption{Example of a single user's VAoI evolution. Black arrows indicate packet generation, red arrows indicate successful packet transmission, and gray dashed arrows indicate failed packet transmission.}
     \label{fig:voi_example}
\vspace{-0.1in}
 \end{figure}
 
Fig. \ref{fig:voi_example} gives an illustrative example for the VAoI evolution for a single user, highlighting how packet generations, successful transmissions, and transmission failures jointly determine the version gap. In the figure, black arrows indicate packet generations that arrive at the scheduler at the beginning of a time slot. Since each scheduled transmission occupies the entire slot, red arrows represent successful deliveries to the destination at the end of the slot, while gray dashed arrows represent failed transmission attempts.

We observe in Fig. \ref{fig:voi_example} that, unlike the classical AoI, which resets based on elapsed time since the generation of the last successful update, VAoI resets only when the version indices at the scheduler and the destination align. At the beginning of slots 1 and 2, new packets with versions 1 and 2 are generated, increasing the VAoI to 1 and 2, respectively. The packet with version 2 is scheduled and successfully delivered. Since no new packet is generated at the beginning of slot 3, the destination version matches the scheduler in slot 3, resetting the VAoI to 0. 

In slot 4, a new packet with version 3 is first generated and then scheduled for transmission, but this attempt fails. With no new generation occurring in slot 5, the VAoI remains at 1. In slot 6, a new packet with version 4 is generated, increasing the VAoI to 2. Although this packet is scheduled for transmission, the delivery again fails, so the destination version remains outdated. In slot 7, the version-4 packet is retransmitted successfully, aligning the scheduler and destination versions and resetting the VAoI to 0 at slot 8 (e.g., no new packet is generated at slot 8). This example underscores that VAoI evolves with version progression and transmission outcomes, rather than merely with the passage of time. As a result, VAoI captures the semantic mismatch between the freshest available update at the scheduler and the most recently delivered update at the destination.

A commonly used risk-neutral performance measure is the long-term system average VAoI, defined as
\begin{align}\label{formula:average_vaoi}
    \bar{\mathcal{D}}=\lim_{T\to\infty}\frac{1}{TN}\sum_{t=1}^{T}
    \sum_{n=1}^{N}\mathcal{D}(t,n),
\end{align}
where $T$ denotes the time horizon. Average VAoI captures typical system behavior but does not account for rare, extreme staleness events.

To characterize tail behavior, such as severe version mismatches caused by bursty arrivals or consecutive transmission failures, we adopt the CVaR at confidence level $\alpha\in(0,1)$. Specifically, we collect all VAoI samples across users and time
\begin{align}
\mathcal{Z}=\{\mathcal{D}(t,n):t=1,\ldots,T,~ n=1,\ldots,N\}
\end{align}
and construct the empirical distribution with the cumulative distribution function (CDF) $F_\mathcal{Z}$. The system CVaR of the mixed empirical distribution is given by
\begin{align}\label{formula:CVaR}
    \mathrm{CVaR}_{\alpha}(\mathcal{Z})=\min_{z\in \mathbb{R}}\left\{z+\frac{1}{(1 -\alpha)TN}\sum_{t=1}^{T}\sum_{n=1}^{N}\big[\mathcal{D}(t,n)-z\big]_{+} \right\},
\end{align}
where $[x]_+ = \max\{x,0\}$ and $z$ denotes the auxiliary Value-at-Risk (VaR) threshold. CVaR quantifies the expected VAoI in the worst $(1-\alpha)$ fraction of system states, providing a meaningful measure of tail risk for real-time status updates.

In many practical applications, the scheduler must maintain fresh information under a transmission cost constraint. The average transmission cost is defined as
\begin{align}\label{formula:cost_constraint}
    \eta=\lim_{T\to\infty}\frac{1}{T}\sum_{t=1}^{T}\mathbb{I}[a_{t}\neq0],
\end{align}
where $\mathbb{I}[\cdot]$ is an indicator function that takes the value 1 if the condition inside holds and 0 otherwise. The scheduling policy must ensure $\eta \leq \eta_{\max}$, where $\eta_{\max} \in (0,1]$ denotes the maximum allowable average transmission cost budget. 

Under the above model and constraint, we develop DRL approaches in Sections \ref{sec:risk_neutral} and \ref{sec:risk_sensitive} tailored to risk-neutral and risk-sensitive VAoI performance objectives, respectively. Section \ref{sec:risk_neutral} presents a deep diffusion model-enabled Soft Actor-Critic (D2SAC) method for minimizing average VAoI, leveraging diffusion models to generate optimal decisions. Section \ref{sec:risk_sensitive} introduces a risk-sensitive distributional version of D2SAC (RS-D3SAC) to minimize CVaR, where distributional DRL is essential because CVaR depends on the entire return distribution, not just its expectation.

\section{Risk-Neutral Scheduling with Deep Diffusion Model-Based SAC}\label{sec:risk_neutral}
In this section, we formulate the risk-neutral VAoI scheduling problem as a constrained Markov decision process (CMDP) and introduce the D2SAC algorithm to learn scheduling policies that minimize the long-term average VAoI under a transmission cost constraint.

\subsection{Risk-Neutral Scheduling Problem Formulation}
In the risk-neutral setting, the objective is to minimize the long-term average system VAoI introduced in \eqref{formula:average_vaoi}, while satisfying the transmission cost constraint in \eqref{formula:cost_constraint}. Let $\{a_{1},\ldots,a_{T}\}$ denote the sequence of scheduling actions over horizon $T$. The optimization problem is expressed as
\begin{align}
\begin{array}{ll}
\displaystyle \min_{a_1,\ldots,a_T} & \bar{\mathcal{D}} \\
\text{s.t.} & \eta \le \eta_{\max},
\end{array}
\end{align}
where $\bar{\mathcal{D}}$ is the long-term average system VAoI, and $\eta$ denotes the long-term average transmission cost. This problem can be represented as a CMDP $(\mathcal{S}, \mathcal{A}, \mathcal{P}, \mathcal{R}, \mathcal{C})$ describing the state space, action space, state transition dynamics, reward function, and cost function, respectively. The system components are defined below.

\textbf{State Space $\mathcal{S}$:} The system state at slot $t$ is the $N$-dimensional VAoI vector
\begin{align}
    s_{t}=\big(\mathcal{D}(t,1),\mathcal{D}(t,2),\ldots,\mathcal{D}(t,N) \big)\in \mathcal{S} ,
\end{align}
where $\mathcal{D}(t,n)\in \{0,1,\ldots,\mathcal{D}_{\max}\}$ is the VAoI of user $n$ at slot $t$. The truncation at $\mathcal{D}_{\max}$ ensures a finite MDP state space. This truncation is consistent with practical systems where data exceeding a freshness threshold becomes obsolete, and also avoids unnecessary analytical complexity.

\textbf{Action Space $\mathcal{A}$:} At each slot, the scheduler selects one user or chooses idleness, $a_t \in \mathcal{A}=\{0,1,\dots,N\}$, where $a_t=0$ means that no user is scheduled, and $a_t=n$ means that user $n$ is selected for transmission. At most one packet may be scheduled per slot.

\textbf{State Transition Dynamics $\mathcal{P}$:} The state transition probability $p(s_{t+1} | s_t,a_t) \in \mathcal{P}$ characterizes the probabilistic evolution of the system state from $s_t$ to $s_{t+1}$ under action $a_t$. In each time slot $t$, user $n$ generates a new status update with probability $r_n$, resulting in a random packet-generation event that determines the freshest version index $G(t,n)$ available at the scheduler. Meanwhile, if user $n$ is scheduled for transmission at slot $t$, the update is successfully delivered with probability $p$, yielding a reception outcome and determining version index $B(t,n)$ at the destination.

As illustrated in Fig.~\ref{fig:voi_example}, the resulting state transition from $s_t$ to $s_{t+1}$ is fully determined by the pair of stochastic processes $\{G(t,n)\}$ and $\{B(t,n)\}$, which are parameterized by the arrival probabilities $r_n$ and the channel success probability $p$. In this work, both $r_n$ and $p$ are assumed to be unknown to the scheduler. Consequently, the state transition dynamics cannot be explicitly characterized, and both the risk-neutral and risk-sensitive VAoI scheduling problems are addressed using model-free reinforcement learning, which learns effective control policies directly through interactions with the environment.

\textbf{Reward Function $\mathcal{R}$ and Cost Function $\mathcal{C}$:} The CMDP is equipped with a reward function $\mathcal{R}: \mathcal{S}\times\mathcal{A}\to\mathbb{R}$, which assigns an immediate reward to guide policy learning
\begin{align}\label{formula:reward_function}
    r_{t}=-\sum_{n=1}^{N}\mathcal{D}(t,n)-\lambda c_{t},
\end{align}
The first term penalizes the system-wide VAoI at slot $t$, encouraging the policy to maintain information freshness. The second term penalizes the instantaneous transmission cost via the Lagrange multiplier $\lambda\ge 0$. The binary cost variable is defined as
\begin{align}
    c_t =
\begin{cases}
1, & a_t \neq 0 \\
0, & a_t = 0
\end{cases}
\end{align}
so that a cost is incurred only when the scheduler transmits a packet. To enforce the long-term constraint $\eta\le\eta_{\max}$, the dual variable $\lambda$ is updated using the projected gradient rule \cite{CMDP}
\begin{align}\label{formula:lambda_updata}
\lambda \leftarrow [\lambda+\delta(\eta-\eta_{\max})]_+,
\end{align}
where $\delta>0$ is a step size and $\eta$ denotes the running average transmission cost. The cumulative return is given by
$R=\sum_{t=0}^{T}\gamma^{t} r_t$,
where $\gamma\in(0,1]$ is a discount factor that weights future rewards relative to immediate ones. Accordingly, we adopt a discounted-return formulation and seek to maximize the expected discounted return under the CMDP framework. This formulation provides a tractable surrogate for the long-term average VAoI minimization problem, while enabling the incorporation of long-term transmission cost constraints.

Despite the conceptual clarity of the CMDP formulation, learning an effective scheduling policy remains challenging due to the intricate VAoI evolution dynamics and the non-stationarity of the reward signal induced by the adaptive dual variable. Moreover, conventional actor-critic architectures often exhibit limited expressive power in representing complex action distributions and may therefore converge to suboptimal scheduling policies. To address these challenges, the next subsection introduces a deep diffusion-model-enhanced Soft Actor-Critic (D2SAC) algorithm, which leverages generative diffusion models to enable richer policy representations and promote more stable learning dynamics.

\begin{figure}[t]
    \centering
    \includegraphics[width=1\linewidth]{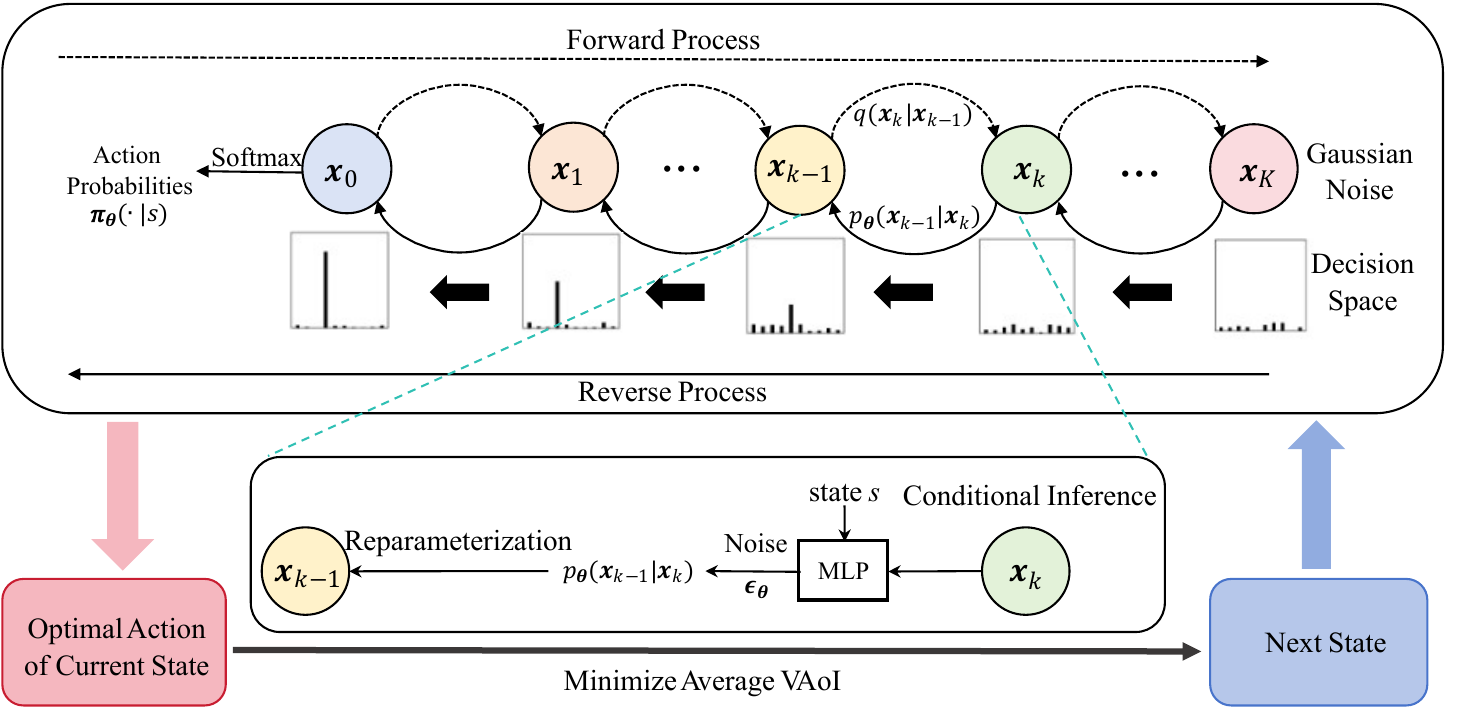}
    \caption{Schematic overview of the diffusion-based policy network.}
    \label{fig:diffuison_detail}
    \vspace{-0.1in}
\end{figure}

\subsection{Deep Diffusion Model-Based SAC}
Diffusion models have recently emerged as powerful generative frameworks for modeling complex, high-dimensional, and multimodal distributions via iterative denoising. Their ability to capture structured dependencies across dimensions makes them well-suited to wireless scheduling problems, where actions often exhibit strong interdependencies. Leveraging these advantages, we incorporate a denoising diffusion probabilistic model (DDPM) into the SAC framework to obtain an expressive and stable policy architecture with improved exploration. The resulting diffusion-based policy network and the overall D2SAC structure are illustrated in Figs. \ref{fig:diffuison_detail} and \ref{fig:diff_sac}, respectively.

\subsubsection{Policy Generation by the Diffusion Model} 
A diffusion model comprises two stochastic components: a \emph{forward} process that gradually corrupts a clean sample with Gaussian noise, and a \emph{reverse} process that reconstructs structured samples from pure noise. We next briefly describe the two processes.

\textbf{Forward Process:} As shown in Fig. \ref{fig:diffuison_detail}, the forward process constructs a Markov chain that iteratively perturbs an initial clean sample $\boldsymbol{x}_0$ drawn from the unknown true distribution into Gaussian noise through $K$ noise-injection steps
\begin{align}
\mathrm{q}(\boldsymbol{x}_{K}|\boldsymbol{x}_{0})
= \prod_{k=1}^{K} \mathrm{q}(\boldsymbol{x}_{k}|\boldsymbol{x}_{k-1}),
\end{align}
where each transition follows
\begin{align}
\mathrm{q}(\boldsymbol{x}_{k}|\boldsymbol{x}_{k-1})
= \mathbf{N}\left(
\boldsymbol{x}_{k};
\sqrt{1-\beta_{k}}\boldsymbol{x}_{k-1},
\beta_{k}\mathbf{I}
\right),
\end{align}
where $\mathbf{N}(\cdot)$ represents Gaussian distribution, $\mathbf{I}$ is the identity matrix, and $\beta_{k}=1-e^{-\frac{\beta_{\min}}{K}-\frac{2k-1}{2K^{2}}\left(\beta_{\max} -\beta_{\min}\right)}$ denotes the forward process variance governed by a variance preserving scheduler \cite{diffusion_DQL}. $\beta_{\min}$ and $\beta_{\max}$ are the minimum and maximum noise levels, respectively. For efficient sampling, the forward process can be expressed in closed form to generate a noisy sample $\boldsymbol{x}_k $ at any arbitrary time step $k$. Expanding the chain yields a closed-form noisy sample at any step $k$
\begin{align}
\mathrm{q}(\boldsymbol{x}_{k}|\boldsymbol{x}_{0})
= \mathbf{N}\left(
\boldsymbol{x}_{k};
\sqrt{\bar{\alpha}_k}\boldsymbol{x}_{0},
(1-\bar{\alpha}_k)\mathbf{I}
\right),
\end{align}
where $\bar{\alpha}_k=\prod_{i=1}^k\alpha_i$ and $\alpha_k=1-\beta_k$.
Thus, sampling can be performed by
\begin{align}
\boldsymbol{x}_k=\sqrt{\bar{\alpha}_k}\boldsymbol{x}_0+\sqrt{1-\bar{\alpha}_k}\boldsymbol{\epsilon},\qquad
\boldsymbol{\epsilon}\sim\mathbf{N}(\mathbf{0},\mathbf{I}). \label{formula:x_k}
\end{align}

In contrast to generative modeling tasks, no dataset of optimal decisions is available in reinforcement learning. Hence, D2SAC does \emph{not} execute the forward process during training; only the reverse process is used to generate policy samples.

\textbf{Reverse Process:}
The reverse diffusion process iteratively reconstructs $\boldsymbol{x}_0$ from noise $\boldsymbol{x}_K\sim\mathbf{N}(\mathbf{0},\mathbf{I})$, as shown in Fig. \ref{fig:diffuison_detail}. In our case, the goal is to recover the latent representation $\boldsymbol{x}_0$ of the optimal decision from a noisy sample, where $\boldsymbol{x}_0=\begin{bmatrix}x_{0}^{0},x_{0}^{1},\cdots,x_{0}^{N}\end{bmatrix}$, which is later mapped into an action distribution. The generative distribution of $\boldsymbol{x}_0$ can be written as
\begin{align}
\mathrm{p}_{\boldsymbol{\theta}}(\boldsymbol{x}_0)=\mathrm{p}(\boldsymbol{x}_K)\prod_{k=1}^{K}\mathrm{p}_{\boldsymbol{\theta}}(\boldsymbol{x}_{k-1}|\boldsymbol{x}_{k}),
\end{align}
where $\mathrm{p}(\boldsymbol{x}_{K})$ is a Gaussian noise distribution and $\boldsymbol \theta$ represents the parameters of the reverse model. The reverse transition is approximated by
\begin{align}\label{formula:reverse_step_distribution}
\mathrm{p}_{\boldsymbol{\theta}}(\boldsymbol{x}_{k-1}|\boldsymbol{x}_{k})=
\mathbf{N}\left(\boldsymbol{x}_{k-1};\boldsymbol{\mu}_{\boldsymbol{\theta}}(\boldsymbol{x}_{k},k,s),\tilde{\beta}_{k}\mathbf{I}\right).
\end{align}
with $\widetilde{\beta}_{k} = \tfrac{1-\bar{\alpha}_{k-1}}{1-\bar{\alpha}_{k}} \beta_{k}$ denoting a deterministically computable variance amplitude. The mean $\boldsymbol{\mu}_{\boldsymbol\theta}$ is parameterized by a neural network conditioned on both the diffusion step index $k$ and the state $s$. Using the forward-process relationship and Bayes' theorem, the reverse mean can be expressed as
\begin{align}\label{eq:mu_from_x0}
{\boldsymbol{\mu}_{\boldsymbol \theta}}(\boldsymbol{x}_{k},k,s)=
\frac{\sqrt{\alpha_{k}}\left(1- \bar{\alpha}_{k-1}\right)}{1-\bar{\alpha}_{k}}
\boldsymbol{x}_{k}
+
\frac{\sqrt{\bar{\alpha}_{k-1}}\beta_{k}}{1-\bar{\alpha}_{k}}
\boldsymbol{x}_{0}.
\end{align}

The reconstruction of original sample $\boldsymbol{x}_0$ can be obtained according to (\ref{formula:x_k})
\begin{align}\label{eq:x0_reconstruction}
\boldsymbol{x}_{0}=
\frac{1}{\sqrt{\bar{\alpha}_{k}}}\boldsymbol{x}_{k}-
\sqrt{
\frac{1}{\bar{\alpha}_{k}}-1
}\cdot
\tanh\left(
\boldsymbol{\epsilon}_{\boldsymbol{\theta}}(\boldsymbol{x}_{k},k,s)
\right),
\end{align}
where $\boldsymbol{\epsilon}_{\boldsymbol{\theta}}$ is the learned denoising noise predictor. The $\tanh(\cdot)$ function here is applied to limit the magnitude of the estimated noise, preventing unstable reconstructions and preserving the structure of the underlying action distribution.

The mean $\boldsymbol{\mu}_\theta$ can thus be written based on (\ref{eq:mu_from_x0}) and (\ref{eq:x0_reconstruction}) 
\begin{align}\label{formula:mu_from_epsilon}
\boldsymbol{\mu}_{\boldsymbol{\theta}}(\boldsymbol{x}_k,k,s)=
\frac{1}{\sqrt{\alpha_k}}
\left(\boldsymbol{x}_k-\frac{\beta_k}{\sqrt{1-\bar{\alpha}_k}}
\tanh\big(\boldsymbol{\epsilon}_{\boldsymbol{\theta}}(\boldsymbol{x}_k,k,s)\big)\right).
\end{align}

To allow gradient backpropagation through stochastic sampling, the reparameterization trick is applied \cite{ddpm}
\begin{align}\label{formula:reverse_sampling}
\boldsymbol{x}_{k-1} = \boldsymbol{\mu}_{\boldsymbol{\theta}}(\boldsymbol{x}_k,k,s)+\sqrt{\tilde{\beta}_k} \boldsymbol{\epsilon},\quad
\boldsymbol{\epsilon}\sim\mathbf{N}(\mathbf{0},\mathbf{I}).
\end{align}
Repeating this denoising process from $k=K$ to $1$ produces the clean sample $\boldsymbol{x}_0$, which is then transformed into an action distribution via the softmax function
\begin{align}\label{formula:softmax}
    \boldsymbol{\pi}_{\boldsymbol{\theta}}\left(\cdot|s\right)=
    \left\{\frac{e^{x_{0}^{i}}}{\sum_{j=0}^{ N}e^{x_{0}^{j}}},\forall i\in\mathcal{A}\right\},
\end{align}
Elements in $\boldsymbol{\pi}_{\boldsymbol{\theta}}(\cdot|s)$ correspond to the probabilities of selecting each possible action in state $s$. In particular, the probability of choosing action $a$ under state $s$ is denoted by $\pi_{\boldsymbol{\theta}}(a|s)$.

\subsubsection{Integrating Diffusion Model Into the SAC Framework}
Soft Actor-Critic (SAC) is a state-of-the-art DRL algorithm grounded in the maximum-entropy RL framework. Its objective is to jointly maximize the expected long-term return and the entropy of the policy. The entropy term encourages the policy to remain sufficiently stochastic, thereby promoting robust and persistent exploration in complex environments. Within SAC, the actor network learns a stochastic policy $\boldsymbol{\pi}_{\boldsymbol{\theta}}(\cdot|s_t)$, while two critic networks estimate the state-action value by minimizing the temporal-difference (TD) error. The double-Q structure mitigates value overestimation and improves stability in policy evaluation and optimization.

\textbf{Diffusion-Based Policy Network:} Building upon the SAC framework, D2SAC replaces the conventional neural policy network (implemented by a multi-layer perceptron) with the diffusion-based policy model illustrated in Fig. \ref{fig:diffuison_detail}. Under the SAC formulation, the global optimization objective remains unchanged, i.e., maximizing the sum of the expected discounted return and the policy entropy. The diffusion model is used to enhance the expressiveness of the policy by generating actions through a multi-step denoising process, enabling the model to capture complex action correlations that frequently arise in multi-user wireless scheduling. 

Formally, the diffusion policy network is trained to determine the optimal parameters $\boldsymbol{\theta}^{*}$ according to
\begin{align}
\boldsymbol{\theta}^{*}
= \arg\max_{\boldsymbol{\theta}}
\mathbb{E}\left[\sum_{t=0}^{T} \gamma^{t}\big(r_{t} + \psi H(\boldsymbol{\pi}_{\boldsymbol{\theta}}(\cdot|s_{t}))\big)\right],
\end{align}
where $\gamma$ is the discount factor and $\psi$ is the temperature coefficient, and $H(\boldsymbol{\pi}_{\boldsymbol{\theta}}(\cdot|s_t)) \triangleq -\boldsymbol{\pi}_{\boldsymbol{\theta}}(\cdot|s_t)^T\log\boldsymbol{\pi}_{\boldsymbol{\theta}}(\cdot|s_t)$ denotes the policy entropy. Equivalently, this maximization can be expressed as a minimization problem
\begin{align}
\boldsymbol{\theta}^{*}
= \arg\min_{\boldsymbol{\theta}}-
\mathbb{E}\left[\sum_{t=0}^{T} \gamma^{t}\big(r_{t} + \psi H(\boldsymbol{\pi}_{\boldsymbol{\theta}}(\cdot|s_{t}))\big)\right],
\end{align}

 \begin{figure}[t]
     \centering
     \includegraphics[width=1\linewidth]{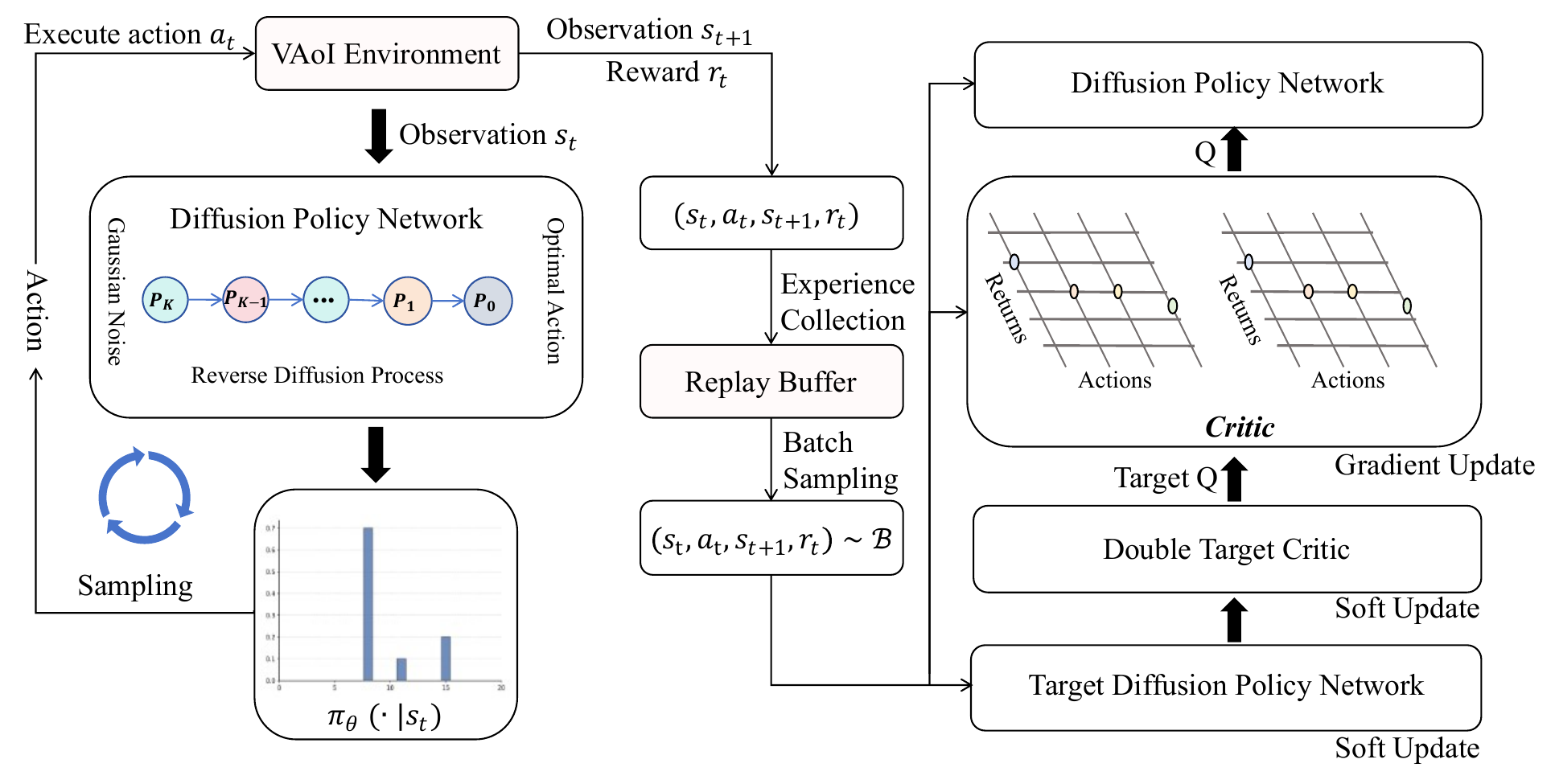}
     \caption{Architecture of the D2SAC algorithm.}
     \label{fig:diff_sac}
     \vspace{-0.1in}
 \end{figure}
 
As shown in Fig. \ref{fig:diff_sac}, the actor in D2SAC is realized by the diffusion policy network $\boldsymbol{\pi}_{\boldsymbol{\theta}}$, whose output distribution is generated through a $K$-step reverse diffusion process. Starting from an initial Gaussian noise vector $\boldsymbol{x}_K\sim\mathbf{N}(\mathbf{0},\mathbf{I})$, the network iteratively denoises the latent representation to obtain a clean sample $\boldsymbol{x}_0$. This final representation is then transformed into an action distribution via the softmax operator in (\ref{formula:softmax}), producing the stochastic policy $\boldsymbol{\pi}_{\boldsymbol{\theta}}(\cdot|s_t)$.

To accommodate this generative policy structure, we denote by a vector-valued Q-function $\boldsymbol{Q}_{\boldsymbol{\phi}}(s_t)$ that includes the Q-values for all actions under state $s_t$, and by $Q_{\boldsymbol{\phi}}(s_t, a_t)$ the scalar Q function for taking action $a_t$ in state $s_t$. The actor's update objective is therefore
\begin{equation}\label{eq:actor_loss}
\min_{\boldsymbol{\theta}}\
-\Big[\boldsymbol{\pi}_{\boldsymbol{\theta}}(\cdot|s_t)^{T} \boldsymbol{Q}_{\boldsymbol{\phi}}({s}_{t})
+\psi H(\boldsymbol{\pi}_{\boldsymbol{\theta}}(\cdot|s_t))\Big],
\end{equation}
where $\boldsymbol{Q}_{\boldsymbol{\phi}}(s_t) = \min\big(\boldsymbol{Q}_{\boldsymbol{\phi}^1}(s_t), \boldsymbol{Q}_{\boldsymbol{\phi}^2}(s_t)\big)$ is the element-wise minimum of the two critic networks with parameters $\boldsymbol{\phi}^1$ and $\boldsymbol{\phi}^2$ (see the details of the critic networks below). This minimum operation is consistent with SAC's conservative value-estimation strategy and effectively suppresses Q-overestimation.

As indicated in Fig. \ref{fig:diff_sac}, a replay buffer $\mathcal{M}$ stores transitions $(s_t,a_t,s_{t+1},r_t)$ collected during agent-environment interaction. Note that the instantaneous cost $c_t$ is used only for online Lagrange multiplier updates, while the replay buffer stores the Lagrangian-shaped reward for policy training. During training, a mini-batch $\mathcal{B}$ of size $b$ transitions is sampled from the replay buffer $\mathcal{M}$. The gradient of the actor loss with respect to $\boldsymbol{\theta}$ is computed as
\begin{align}\label{formula:actor_grad}
\nabla_{\boldsymbol{\theta}}\mathcal{L}(\boldsymbol{\theta})
= \mathbb{E}_{s_t \sim \mathcal{B}}\Big[
    & - \nabla_{\boldsymbol{\theta}}
      \boldsymbol{\pi}_{\boldsymbol{\theta}}(\cdot|s_t)^{T} 
      \boldsymbol{Q}_{\boldsymbol{\phi}}(s_t) \notag \\
    & ~~~~~ ~~~~~-\psi \nabla_{\boldsymbol{\theta}}
      H\!\left(\boldsymbol{\pi}_{\boldsymbol{\theta}}(\cdot|s_t)\right)
\Big].
\end{align}
where $\boldsymbol\theta$ and $\boldsymbol\phi$ denote the policy and Q-function parameters, respectively. The first term drives the policy toward actions with high estimated value, while the second term promotes exploration by encouraging higher entropy (i.e., serving as an action entropy regularizer). The policy parameters are then updated via gradient descent based on (\ref{formula:actor_grad})
\begin{align}\label{formula:gradient_update_theta}
    \boldsymbol{\theta}\leftarrow\boldsymbol{\theta}-{\omega_{ {a}}}\cdot\nabla_{\boldsymbol{\theta}}\mathcal{L}\big( \boldsymbol{\theta}\big),
\end{align}
where $\omega_{{a}}$ is the actor's learning rate. To stabilize training and avoid oscillations, a slowly-updated target policy network with parameters $\boldsymbol{\hat{\theta}}$ is maintained. Its parameters are updated periodically using a soft update mechanism
\begin{align}\label{for:actor_soft_update}
\begin{split}
\boldsymbol{\hat{\theta}}&
\leftarrow\zeta\boldsymbol{\theta}+(1-\zeta)\boldsymbol{\hat{ \theta}},
\end{split}
\end{align}
where $\zeta \in (0,1)$ is the soft update coefficient, which controls the stability of the target network.

\textbf{Double Critic Network:} In the D2SAC architecture (Fig. \ref{fig:diff_sac}), the critic networks aim to estimate the Q-values with high accuracy. To enhance robustness, SAC adopts a double-critic architecture with parameters $\boldsymbol{\phi}^{1}$ and $\boldsymbol{\phi}^{2}$. Each critic is updated independently and outputs a vector $\boldsymbol{Q}_{\boldsymbol{\phi}^i}(s_t)$, providing Q-values for all actions in state $s_t$. Consistent with the actor, each critic has a corresponding target network, with parameters $\hat{\boldsymbol{\phi}^1}$ and $\hat{\boldsymbol{\phi}^2}$.

During policy evaluation, the smaller output of the two Q-networks is used to construct the target, mitigating overestimation bias and improving convergence. The target value is
\begin{align}
\hat{y}=r_t+\gamma\Big(
\boldsymbol{\pi}_{\hat{\boldsymbol{\theta}}}(\cdot|s_{t+1})^{T}
\min_{i=1,2} \boldsymbol{Q}_{\hat{\boldsymbol{\phi}^i}}(s_{t+1})+\psi H(\boldsymbol{\pi}_{\hat{\boldsymbol{\theta}}}(\cdot|s_{t+1}))
\Big),
\end{align}
where $\boldsymbol{\pi}_{\hat{\boldsymbol{\theta}}}$ and $Q_{\hat{\boldsymbol{\phi}^i}}$ denote the outputs of the target actor and the two target critic networks. The current Q-value generated by critic $i$ is
\begin{align}
y^{i}=Q_{{\boldsymbol{\phi}}^{i}}\left(s_{t},a_{t}\right)
\end{align}
where $Q_{{\boldsymbol{\phi}}^{i}}(s_{t},a_{t})$ denotes the Q-value corresponding to the action $a_t$ produced by the $i$-th critic $\boldsymbol{Q}_{\boldsymbol{\phi}^{i}}(s_t)$. The critic parameters are updated by minimizing the TD error between $\hat{y}$ and $y^i$
\begin{align}\label{formula:double_q_loss}
    \min_{\boldsymbol{\phi}^{1},\boldsymbol{\phi}^{2}} \mathbb{E}_{(s_{t},a_{t},s_{t+1},r_{t} )\sim\mathcal{B}}\left[\sum_{i=1,2}\left(\hat{y}-y^{i}\right)^{2}\right].
\end{align}

By minimizing the TD error in (\ref{formula:double_q_loss}), we can continuously improve the estimation accuracy of the critics, thereby providing more reliable Q-value evaluations for policy improvement. As in the target actor network, the parameters of the target critics are updated using a soft update mechanism
\begin{equation}\label{formula:critic_soft_update}
    \hat{\boldsymbol{\phi}^i} \leftarrow \zeta \boldsymbol{\phi}^i + (1-\zeta)\hat{\boldsymbol{\phi}^i},\; i=1,2.
\end{equation}

\begin{algorithm}[t]
\caption{D2SAC: Deep Diffusion Soft Actor-Critic}
\label{alg:diffusion_sac}
\begin{algorithmic}[1]

\STATE \textbf{Initialize:} Replay buffer $\mathcal{M}$; diffusion actor network with parameters $\boldsymbol{\theta}$ and target actor with parameters $\hat{\boldsymbol{\theta}}$; two critic networks with parameters $\boldsymbol{\phi}^1$ and $\boldsymbol{\phi}^2$, and corresponding target critic networks with parameters $\hat{\boldsymbol{\phi}^1}$ and $\hat{\boldsymbol{\phi}^2}$; Lagrange multiplier $\lambda \ge 0$; empirical average cost estimate $\eta = 0$; global step counter $l =0$.

\FOR{each training iteration}

    \STATE Observe current state $s_1$    
    \FOR{environment step $t = 1$ to $T$}    
        \STATE Sample initial diffusion noise $\boldsymbol{x}_K \sim \mathbf{N}(\mathbf{0}, \mathbf{I})$
        
        \FOR{denoising step $k = K$ down to $1$}
            \STATE Compute scaled denoising prediction             $\tanh\big(\boldsymbol{\epsilon}_{\theta}(\boldsymbol{x}_{k}, k, s_t)\big)$
            using the diffusion actor network
            \STATE Compute reverse-process mean $\boldsymbol{\mu}_{\boldsymbol{\theta}}$ using (\ref{formula:reverse_step_distribution})-(\ref{formula:mu_from_epsilon})\STATE Sample $\boldsymbol{x}_{k-1}$ via reparameterization (\ref{formula:reverse_sampling})
        \ENDFOR
        
        \STATE Obtain action distribution from $\boldsymbol{x}_0$ using (\ref{formula:softmax}) and sample action $a_t \sim \boldsymbol{\pi}_{\boldsymbol{\theta}}(\cdot|s_t)$        
        \STATE Execute $a_t$, observe next state $s_{t+1}$, reward $r_t$, and instantaneous cost $c_t$
        \STATE Store transition $(s_t, a_t, s_{t+1},  r_t)$ in replay buffer $\mathcal{M}$ \STATE $l \leftarrow l + 1$ \STATE Update empirical average cost: $\eta \leftarrow \eta + \frac{1}{l}(c_t - \eta)$
        \STATE Set $s_t \leftarrow s_{t+1}$
    \ENDFOR
    
   \STATE Update Lagrange multiplier $\lambda$ according to (\ref{formula:lambda_updata})
    
    \STATE Sample minibatch $\mathcal{B} \subset \mathcal{M}$
\STATE Update critic parameters $\boldsymbol{\phi}^1$, $\boldsymbol{\phi}^2$ by minimizing the double-Q loss in (\ref{formula:double_q_loss})
    \STATE Update diffusion actor parameters $\boldsymbol{\theta}$ using the policy gradient method (\ref{formula:actor_grad}) and (\ref{formula:gradient_update_theta})
    
    \STATE Soft-update target networks by (\ref{for:actor_soft_update}) and (\ref{formula:critic_soft_update})
\ENDFOR

\RETURN Well-trained D2SAC policy $\boldsymbol{\pi}_{\boldsymbol{\theta}^*}$

\end{algorithmic}
\end{algorithm}

Algorithm~\ref{alg:diffusion_sac} summarizes the overall D2SAC training pipeline that minimizes the average VAoI under a transmission cost constraint. In classical stochastic primal-dual algorithms for CMDPs, the primal variables (actor-critic parameters) are typically updated on a faster timescale than the dual variable $\lambda$. In Algorithm~\ref{alg:diffusion_sac}, we adopt a simplified implementation in which $\lambda$ and the actor-critic parameters are each updated once per training iteration. Importantly, the dual update relies on a running average of accumulated constraint costs, whereas the actor-critic parameters are updated using stochastic minibatch gradients drawn from the replay buffer. As a result, the dual variable evolves in response to a more slowly varying signal, leading to an effective separation of timescales in practice. The framework naturally extends to multiple primal updates per dual update, consistent with standard two-timescale stochastic optimization. By integrating a diffusion-based generative policy with the maximum-entropy RL framework, D2SAC achieves strong policy expressiveness and improved action modeling capability, thanks to the multi-step denoising process. 

However, like standard actor-critic methods, D2SAC is inherently risk-neutral, optimizing only the expected information freshness. In practical wireless systems, VAoI can occasionally spike due to bursty arrivals or consecutive transmission failures, leading to severe semantic staleness that threatens reliability in latency-sensitive or safety-critical applications. To address these rare but severe events, a risk-sensitive formulation is necessary. The following section presents a distributional extension of D2SAC that models the full return distribution and optimizes policies under CVaR-based criteria.

\section{Risk-Sensitive Scheduling with Deep Distributional Diffusion Model-Based SAC}\label{sec:risk_sensitive}
This section develops a risk-sensitive extension of D2SAC tailored for multi-user VAoI scheduling. We first formulate the scheduling problem under the CVaR criterion, and then introduce the RS-D3SAC algorithm, which integrates a diffusion-based generative actor with a distributional critic to learn return distributions and enable CVaR-driven tail-aware control.

\subsection{Risk-Sensitive Scheduling Problem Formulation}
To explicitly capture the extreme semantic staleness of the system-wide
VAoI, we adopt the CVaR metric in (\ref{formula:CVaR}), which evaluates the expected VAoI in the worst $(1-\alpha)$ fraction of system realizations. Incorporating the long-term transmission cost constraint in (\ref{formula:cost_constraint}), the resulting risk-sensitive scheduling problem is formulated as
\begin{align}
\begin{array}{ll}
\displaystyle \min_{a_1,\ldots,a_T} & \mathrm{CVaR}_{\alpha}(\mathcal{Z}) \\
\text{s.t.} & \eta \le \eta_{\max},
\end{array}
\end{align}
The CMDP structure, including the state and action spaces, transition dynamics, reward function, and transmission cost, is assumed identical to the risk-neutral formulation in Section \ref{sec:risk_neutral}. This is an assumption for consistency: although one could design a new reward function for risk-sensitive control, our goal is to retain the original formulation that achieves good average VAoI performance. Building upon this baseline, we augment the learning objective to account for tail risks using CVaR, thereby improving robustness to extreme staleness events without compromising the average-case behavior.

Unlike expectation-based DRL methods that consider only the mean return, the proposed RS-D3SAC algorithm models the full return distribution through quantile-based value estimation. This enables explicit CVaR evaluation and tail-aware updates, providing a principled mechanism to mitigate extreme semantic staleness in VAoI, as detailed in subsequent sections. 

\subsection{D3SAC: Distributional D2SAC}
In D2SAC, each critic outputs a vector-valued Q-function $\boldsymbol{Q}_{\boldsymbol{\phi}}(s_t)$ (we omit the index of the critic to improve clarity), where each scalar component $Q_{\boldsymbol{\phi}}(s_t, a_t)$ represents the expected discounted return for taking action $a_t$ in state $s_t$. While this scalar expectation is adequate for optimizing long-term average performance, it provides no information about the variability of returns or the likelihood of extreme tail events. As a result, risk-neutral D2SAC cannot differentiate between actions with similar mean returns but vastly different risk characteristics. To address this limitation, D3SAC replaces the scalar critic with a distributional quantile critic capable of modeling the entire return distribution. While the actor remains to employ the diffusion-based generative model used in D2SAC, the primary modification lies in redesigning the critic to capture uncertainty and tail behavior, thereby enabling risk-sensitive decision making.

\begin{figure}[t]
    \centering
    \includegraphics[width=1\linewidth]{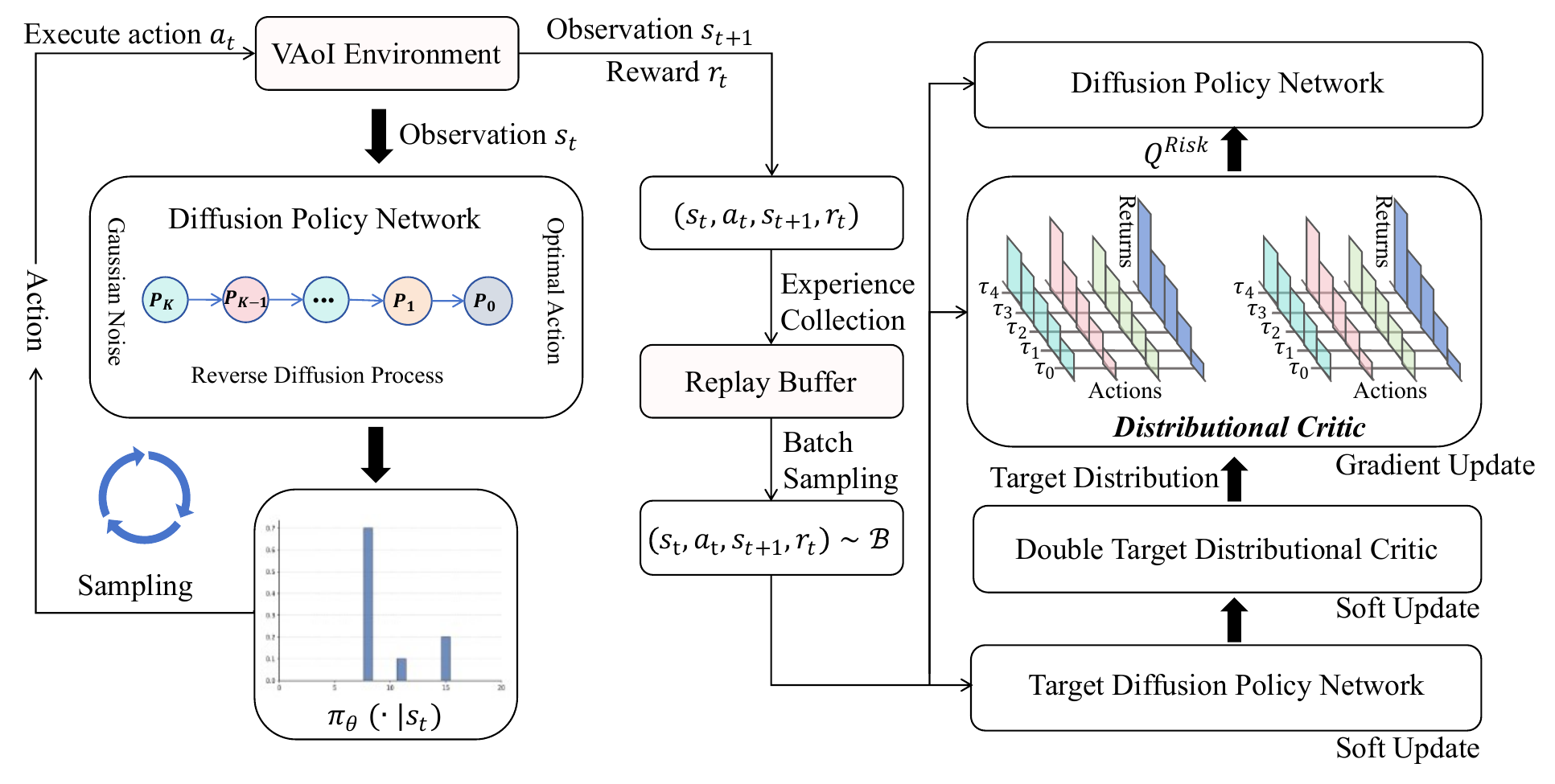}
    \caption{Architecture of the proposed RS-D3SAC algorithm.}
    \label{fig:RS-D3SAC}
\end{figure}

Instead of estimating the scalar value $Q_{\boldsymbol{\phi}}(s_t, a_t)$, D3SAC employs distributional RL to model the return as a random variable, $Z_{\boldsymbol{\xi}}(s_t,a_t)$, where $\boldsymbol{\xi}$ denotes the parameters of the distributional critic. The conventional Q-value is recovered as the mean of the learned distribution, $Q_{\boldsymbol{\xi}}(s_t, a_t)=\mathbb{E}[Z_{\boldsymbol{\xi}}(s_t,a_t)]$. By learning the full distribution, the policy becomes sensitive to rare but severe VAoI spikes, which often dominate performance in real-time or safety-critical applications.

The distributional critic satisfies the distributional Bellman equation \cite{qrdqn}
\begin{align}
Z_{\boldsymbol{\xi}}(s_t,a_t)
\overset{\mathbf D}{=}
r_t+ \gamma\Big(
Z_{\boldsymbol{\xi}}(s_{t+1}, a')
-
\psi \log \pi_{\hat{\boldsymbol{\theta}}}(a'|s_{t+1})
\Big),
\end{align}
where $\overset{\mathbf{D}}{=}$ denotes equality in distribution, and $Z_{\boldsymbol{\xi}}(s_{t+1},a')$ denotes the distribution of future returns under action $a'$ in the next state $s_{t+1}$, $a' \sim \boldsymbol{\pi}_{\hat{\boldsymbol{\theta}}}(\cdot|s_{t+1})$. Modeling $Z_{\boldsymbol{\xi}}(s_t,a_t)$ directly, rather than its expectation, enables the algorithm to penalize low-quantile (i.e., worst-case) returns associated with large VAoI realizations. 

To approximate the return distribution, D3SAC adopts the Quantile Regression DQN (QR-DQN) representation. It approximates the CDF of the return by learning a set of quantiles. Let the cumulative probability of $Z_{\boldsymbol{\xi}}(s_t,a_t)$ be divided into $\mathcal{N}$ uniform probability intervals $\{\tau_1, \tau_2,..., \tau_j,..., \tau_\mathcal{N}\}$, where $\tau_j=j/\mathcal{N}$. For each interval $(\tau_{j-1}, \tau_j)$, QR-DQN defines the midpoint $\hat \tau_j=(\tau_{j-1}+\tau_j)/2,~j=\{1,2,...,\mathcal{N}\}$. The critic then learns the corresponding quantile values $\sigma_j(s_t,a_t;\boldsymbol{\xi})$, representing the $\hat{\tau}_j$-quantile of the return distribution. Under this representation, the distribution $Z_{\boldsymbol{\xi}}(s_t,a_t)$ is approximated by a uniform mixture of Dirac masses at these learned quantiles
\begin{align}
Z_{\boldsymbol{\xi}}(s_t,a_t)
= \frac{1}{\mathcal{N}} \sum_{j=1}^{\mathcal{N}}
\delta_{\sigma_j(s_t,a_t;\boldsymbol{\xi})},
\end{align}
where $\delta_z$ denotes a Dirac distribution centered at $z$. This representation provides a flexible approximation suitable for modeling the skewed or heavy-tailed behavior often observed in semantic freshness metrics such as VAoI.

As in D2SAC, D3SAC employs a double-critic architecture to improve stability and reduce overestimation bias. Two independent distributional critics, parameterized by $\boldsymbol{\xi}^{1}$ and $\boldsymbol{\xi}^{2}$, generate independent estimates $Z_{\boldsymbol{\xi}^{1}}(s_t,a_t)$ and $Z_{\boldsymbol{\xi}^{2}}(s_t,a_t)$. Corresponding target networks are denoted by $\hat{\boldsymbol{\xi}^{1}}$ and $\hat{\boldsymbol{\xi}^{2}}$.

Given the next state $s_{t+1}$, the target action $a'$ is obtained using the target diffusion-based actor network $\boldsymbol{\pi}_{\hat{\boldsymbol{\theta}}}(\cdot|s_{t+1})$ (as described in Section \ref{sec:risk_neutral}). The two target critics then produce the next-state return distributions $Z_{\hat{\boldsymbol{\xi}^1}}(s_{t+1},a')$ and $Z_{\hat{\boldsymbol{\xi}^2}}(s_{t+1},a')$. Following the double-critic principle, the target distribution, denoted by $Z_{\hat{\boldsymbol{\xi}}}(s_{t+1},a')$, is obtained by taking the element-wise minimum of the quantile outputs
\begin{align}
Z_{\hat{\boldsymbol{\xi}}}(s_{t+1},a')
&= \frac{1}{\mathcal{N}} \sum_{j=1}^{\mathcal{N}}
\delta_{\sigma_j(s_{t+1},a';\hat{\boldsymbol{\xi}})}  \notag \\
&= \frac{1}{\mathcal{N}} \sum_{j=1}^{\mathcal{N}}
\delta_{\min \{ \sigma_j(s_{t+1},a';\hat{\boldsymbol{\xi}^1}),\sigma_j(s_{t+1},a';\hat{\boldsymbol{\xi}^2})\}}
\end{align}

Each distributional critic is trained by minimizing the quantile regression Huber loss across all quantile pairs
\begin{align}
    \mathcal{L}_{}(\boldsymbol{\xi}^{i})=\frac{1}{\mathcal{N}}\sum_{j=1}^{\mathcal{N}}\sum_{j'=1}^{\mathcal{N}}\rho_{ \hat{\tau_{j}}}^\kappa\big(\Delta_{jj^{\prime}}^i\big),\quad i=1,2,
\end{align}
where $\Delta_{jj'}^i$ is the TD error corresponding to the quantile estimates, given by 
\begin{align}
\Delta_{jj'}^{i}=r_t&+\gamma \Big(
\sigma_{j'}(s_{t+1}, a'; \hat{\boldsymbol{\xi}})
- \psi \log \pi_{\hat{\boldsymbol{\theta}}}(a'|s_{t+1})
\Big) \notag \\
&-\sigma_j(s_t, a_t; \boldsymbol{\xi}^{i}),
\quad i=1,2
\end{align}
where $a' \sim \boldsymbol{\pi}_{\hat{\boldsymbol{\theta}}}(\cdot|s_{t+1})$ and $\sigma_j(\cdot)$ denotes the $j$-th quantile of the return distribution. The quantile Huber loss $\rho_{\tau}^{\kappa}(\cdot)$ is defined by 
\begin{align}
    \rho_{\tau}^{\kappa}(u)=\begin{cases}|\tau-\mathbb{I}\{u<0\}|\cdot \frac{1}{2}u^{2},& |u|\leq\kappa,\\ |\tau-\mathbb{I}\{u<0\}|\cdot\kappa(|u|-\frac{1}{2}\kappa),&\mathrm{otherwise},\end{cases}
\end{align}
where $\mathbb{I}\{\cdot\}$ is the indicator function and $\kappa$ denotes the transition threshold between quadratic and linear loss regimes. 

The total loss used for training is the sum of the two distributional critic losses
\begin{align}\label{formula:double_dis_loss}
    \mathcal{L}_{total}=\mathcal{L}_{}(\boldsymbol{\xi}^{1})+\mathcal{L}_{}(\boldsymbol{\xi}^{2}).
\end{align}
By minimizing this objective, the two distributional critics collectively approximate the full return distribution, enabling D3SAC to capture both central tendencies and tail risks. After each update, the target distributional critic parameters are softly updated according to
\begin{align}
\hat{\boldsymbol{\xi}^i} \leftarrow \zeta \boldsymbol{\xi}^i + (1-\zeta)\hat{\boldsymbol{\xi}^i},\; i=1,2.
\label{formula:critic_soft_update_dist}
\end{align}

\begin{algorithm}[t]
\caption{RS-D3SAC: Risk-Sensitive Deep Distributional Diffusion Soft Actor-Critic}
\label{alg:distributional_diff_sac}
\begin{algorithmic}[1]

\STATE \textbf{Initialize:} Replay buffer $\mathcal{M}$; diffusion actor network with parameters $\boldsymbol{\theta}$ and target actor with parameters $\hat{\boldsymbol{\theta}}$; two distributional critic networks with parameters $\boldsymbol{\xi}^1$ and $\boldsymbol{\xi}^2$, and corresponding target critic networks with parameters $\hat{\boldsymbol{\xi}^1}$ and $\hat{\boldsymbol{\xi}^2}$; Lagrange multiplier $\lambda \ge 0$; empirical average cost estimate $\eta = 0$; global step counter $l =0$; number of quantile samples $\mathcal{N}$; CVaR confidence level $\alpha~(\varphi=1-\alpha)$. 
\label{line:dis_ini}

\FOR{each training iteration}
    \STATE Observe current state $s_1$

    \FOR{environment step $t = 1$ to $T$}  
        \STATE Sample initial diffusion noise $\boldsymbol{x}_K \sim \mathbf{N}(\mathbf{0},\mathbf{I})$

        \FOR{denoising step $k = K$ down to $1$}
            \STATE Compute scaled denoising prediction             $\tanh\!\big(\boldsymbol{\epsilon}_{\theta}(\boldsymbol{x}_{k}, k, s_t)\big)$
            using the diffusion actor network
            \STATE Compute reverse-process mean $\boldsymbol{\mu}_{\boldsymbol{\theta}}$ using (\ref{formula:reverse_step_distribution})-(\ref{formula:mu_from_epsilon})\STATE Sample $\boldsymbol{x}_{k-1}$ via reparameterization (\ref{formula:reverse_sampling})
        \ENDFOR

        \STATE Obtain action distribution from $\boldsymbol{x}_0$ using (\ref{formula:softmax}) and sample action $a_t \sim \boldsymbol{\pi}_{\boldsymbol{\theta}}(\cdot|s_t)$        
        \STATE Execute $a_t$, observe next state $s_{t+1}$, reward $r_t$, and instantaneous cost $c_t$
        \STATE Store transition $(s_t, a_t, s_{t+1}, r_t)$ in replay buffer $\mathcal{M}$ \STATE $l \leftarrow l + 1$ \STATE Update empirical average cost: $\eta \leftarrow \eta + \frac{1}{l}(c_t - \eta)$
        \STATE Set $s_t \leftarrow s_{t+1}$
    \ENDFOR

    \STATE Update Lagrange multiplier $\lambda$ according to (\ref{formula:lambda_updata})

    \STATE Sample minibatch $\mathcal{B} \subset \mathcal{M}$

    \STATE Update distributional critics $\boldsymbol{\xi}^1, \boldsymbol{\xi}^2$
    by minimizing the quantile Huber loss (\ref{formula:double_dis_loss}) \label{line:update_dis}

    \STATE Compute the risk-sensitive vector-valued Q-function  $\boldsymbol{Q}_{\boldsymbol{\xi}}^{\text{Risk}}$, where each CVaR-based  Q-value   $Q_{\boldsymbol{\xi}}^{\text{Risk}}$ is obtained by (\ref{formula:Risk-Q}) \label{line:obtain_RS_Q}
   \STATE Update diffusion actor parameters $\boldsymbol{\theta}$ using the policy gradient method (\ref{formula:actor_grad}) and (\ref{formula:gradient_update_theta}) by replacing $\boldsymbol{Q}_\phi$ with $\boldsymbol{Q}_{\boldsymbol{\xi}}^\text{Risk}$\label{line:use_RS_Q}

    \STATE Soft-update target networks using
    (\ref{for:actor_soft_update}) and (\ref{formula:critic_soft_update_dist})

\ENDFOR

\RETURN Well-trained RS-D3SAC policy $\boldsymbol{\pi}_{\boldsymbol{\theta}^*}$

\end{algorithmic}
\end{algorithm}

\subsection{RS-D3SAC for Risk-Sensitive VAoI Scheduling}
We now propose risk-sensitive D3SAC (RS-D3SAC) for CVaR-based VAoI scheduling. Although our ultimate objective is to minimize the CVaR of the system-wide mixed VAoI distribution as expressed in (\ref{formula:CVaR}), directly optimizing this target is computationally challenging. This is because evaluating the CVaR of the mixed VAoI distribution requires analyzing the joint tail distribution of multiple interdependent random variables from multiple users, which is computationally prohibitive in high-dimensional state spaces of our multi-user scenarios. To obtain a tractable surrogate, we leverage the learned return distribution $Z_{\boldsymbol{\xi}}$ to reduce risk.

For a risk level $\varphi\in[0,1]$, the VaR and CVaR of the return distribution $Z_{\boldsymbol{\xi}}$ are defined as
\begin{align}
\mathrm{VaR}_{\varphi}[Z_{\boldsymbol{\xi}}]
&= \inf\{z \in \mathbb{R} : F_{Z_{\boldsymbol{\xi}}}(z) \ge \varphi\}, \\
\mathrm{CVaR}_{\varphi}[Z_{\boldsymbol{\xi}}]
&= \mathbb{E}\left[z \mid z \le \mathrm{VaR}_{\varphi}[Z_{\boldsymbol{\xi}}]\right],
\end{align}
where $F_Z(z)$ is the CDF of the return distribution. With distributional RL, the CVaR is readily computed by averaging quantile samples with $\hat{\tau}_j \le \varphi$. We therefore optimize $\mathrm{CVaR}_{\varphi}[Z_{\boldsymbol{\xi}}]$ as a practical risk-sensitive surrogate.

Since the one-step reward in (\ref{formula:reward_function}) equals the negative VAoI minus transmission cost, the lower tail of the return distribution corresponds to the upper tail of the VAoI distribution. Consequently, maximizing the CVaR of the return effectively reduces high-VAoI outcomes, providing control over the worst-case VAoI. Formally, we define the risk-sensitive Q-value as
\begin{align}
\label{formula:Risk-Q}
    Q_{\boldsymbol{\xi}}^\text{Risk}({s_t,a_t})=\mathrm{CVaR}_{\varphi}[Z_{\boldsymbol \xi}(s_t,a_t)].
\end{align}
Given this definition, let us denote the risk-sensitive vector-valued Q-function $\boldsymbol{Q}_{\boldsymbol{\xi}}^\text{Risk}(s_t)$ that includes the risk-sensitive Q-values for all actions under state $s_t$. The risk-sensitive actor update objective becomes
\begin{align}
\min_{\boldsymbol{\theta}}\
-\Big[\boldsymbol{\pi}_{\boldsymbol{\theta}}(\cdot|s_t)^{T} \boldsymbol{Q}_{\boldsymbol{\xi}}^\text{Risk}({s}_{t})
+\psi H(\boldsymbol{\pi}_{\boldsymbol{\theta}}(\cdot|s_t))\Big],
\end{align}
while the remaining steps of the actor training pipeline remain identical to those in Section~\ref{sec:risk_neutral}, excepting that $\boldsymbol{Q}_{\boldsymbol{\phi}}$ is replaced by $\boldsymbol{Q}_{\boldsymbol{\xi}}^\text{Risk}$. The formulation of RS-D3SAC shifts the policy from expectation-oriented to tail-risk-oriented, aligning learning to mitigate extreme staleness events.

In practice, we set $\varphi = 1 - \alpha$ to remain consistent with the definition of system CVaR of VAoI at confidence level $\alpha$, i.e., $\mathrm{CVaR}_{\alpha}(\mathcal{Z})$. Algorithm~\ref{alg:distributional_diff_sac} summarizes the complete RS-D3SAC procedure. Compared to D2SAC in Algorithm~\ref{alg:diffusion_sac}, the main differences in Algorithm~\ref{alg:distributional_diff_sac} are
\begin{itemize}
    \item[(1)] Initialization of the number of quantiles $\mathcal{N}$ and the CVaR confidence level $\alpha$ (Line~\ref{line:dis_ini}).
    \item[(2)] Replacement of D2SAC's scalar critics with double distributional critics trained using the quantile Huber loss (Line~\ref{line:update_dis}).
    \item[(3)] Computation of CVaR-based risk-sensitive Q-values to guide policy improvement (Lines~\ref{line:obtain_RS_Q} and \ref{line:use_RS_Q}), allowing the actor to prioritize worst-case VAoI reduction.
\end{itemize}

Compared with D2SAC, RS-D3SAC introduces additional computational overhead only in the critic update. Specifically, with $\mathcal{N}$ quantile outputs, the forward and backward passes of the critic scale linearly with $\mathcal{N}$, while the quantile Huber loss incurs an additional $\mathcal{O}(\mathcal{N}^2)$ cost due to pairwise quantile comparisons. The CVaR computation further introduces an $\mathcal{O}(\mathcal{N}\log \mathcal{N})$ overhead to identifying the worst $(1-\alpha)\mathcal{N}$ quantiles via sorting and partial selection. In practice, $\mathcal{N}$ is chosen to be moderate (e.g., $\mathcal{N}=32$ or $64$), and these costs are negligible compared to the diffusion-based actor, which dominates the overall runtime through $K$ denoising steps. Therefore, RS-D3SAC achieves improved tail-risk control with only a modest increase in computational complexity. As demonstrated next, RS-D3SAC significantly improves both average performance and CVaR performance, enabling the system to maintain semantic freshness effectively.

\section{Performance Evaluation} \label{sec:Performance Evaluation}
This section comprehensively evaluates the proposed D2SAC and RS-D3SAC algorithms in terms of both the average VAoI and the CVaR of VAoI. Through extensive simulations, we demonstrate that the proposed methods consistently outperform existing DRL baselines. Moreover, we isolate and examine the individual contributions of the diffusion-based actor and the distributional critic, thereby showing how each architectural component impacts performance under both average and tail-risk criteria.

\subsection{Simulation Setup}
All implementations and simulations are carried out using PyTorch and executed on a high-performance computing server equipped with an Intel Xeon Gold 6226R CPU, an NVIDIA GeForce RTX 4090 GPU, and 512 GB of RAM, running Ubuntu Linux. This computational setup provides sufficient capacity to support the training of diffusion-based policies and distributional value estimators. We implemented two diffusion-based DRL algorithms: D2SAC and its risk-sensitive extension RS-D3SAC. The detailed configurations of the actor, scalar critic, and distributional critic networks are summarized in Table~\ref{tab:actor_critic_networks}.

In both algorithms, the actor is implemented as a diffusion-based generative policy network. The actor network implements a time-conditioned denoising policy that maps Gaussian noise and the current system state to a denoised action distribution. At each diffusion step, a sinusoidal positional embedding encodes the denoising timestep. It is integrated into the actor network to condition its computation on the current stage of the diffusion process. The actor backbone consists of three fully connected layers with Mish activations, followed by a final layer with a Tanh activation to constrain the outputs within a bounded range. During policy execution, the same time-conditioned actor network is applied iteratively for $K$ denoising steps, with each forward pass corresponding to one denoising operation that progressively refines the action.

\begin{table}[t]
    \centering
    \caption{Network Architectures of the Actor, Scalar Critic, and Distributional Critic with $N$ users}
    \label{tab:actor_critic_networks}
    \begin{tabular}{c|c|c|c}
        \hline
        \textbf{Networks} & \textbf{Layer} & \textbf{Activation} & \textbf{Units} \\ 
        \hline
        \multirow{7}{*}{Actor} 
            & SinusoidalPosEmb & -- & 16 \\ 
            & FullyConnect & Mish & 32 \\ 
            & FullyConnect & -- & 16 \\ 
            & Concatenation & -- & -- \\ 
            & FullyConnect & Mish & 256 \\ 
            & FullyConnect & Mish & 256 \\ 
            & FullyConnect & Tanh & $N+1$ \\ 
        \hline
        \multirow{3}{*}{Scalar Critic} 
            & FullyConnect & Mish & 256 \\ 
            & FullyConnect & Mish & 256 \\ 
            & FullyConnect & -- & $N+1$ \\ 
        \hline
        \multirow{3}{*}{Distributional Critic} 
            & FullyConnect & Mish & 256 \\ 
            & FullyConnect & Mish & 256 \\ 
            & FullyConnect & -- & $(N+1)\times \mathcal{N}$ \\ 
        \hline
    \end{tabular}
    \vspace{-0.05in}
\end{table}

\begin{table}[t]
\centering
\caption{Training Parameters.}
\label{tab:parameters}
\begin{tabular}{clc}
\hline
\textbf{Symbol} & \textbf{Description} & \textbf{Value} \\
\hline
$\omega_a$ & Learning rate of the actor network & $2 \times 10^{-4}$ \\
$\omega_c$ & Learning rate of the critic network & $2 \times 10^{-3}$ \\
$\psi$ & Temperature of action entropy regularization & $0.05$ \\
$\zeta$ & Soft update coefficient & $0.005$ \\
$b$ & Batch size of $\mathcal{B}$ & 512 \\
$\gamma$ & Discount factor & $0.95$ \\
$K$ & Number of diffusion denoising steps & 5 \\
$|\mathcal{M}|$ & Replay buffer capacity & $5 \times 10^{6}$ \\
$T$ & Transitions per training iteration & 1000 \\
$\kappa$ & Huber loss threshold & 1 \\
$\delta$ & Lagrange update step size & 1 \\
$\mathcal{N}$ & Number of quantiles in distributional critic & 64 \\
$\alpha$ & CVaR confidence level ($\varphi=1-\alpha$) & 0.75 \\
\hline
\end{tabular}
\vspace{-0.1in}
\end{table}

D2SAC employs a double-critic architecture with two homogeneous scalar critics to mitigate value overestimation. Each critic consists of three fully connected layers with Mish activations followed by a scalar output layer that estimates the state-action value. In RS-D3SAC, the distributional critics share the same backbone architecture but replace the scalar output with multiple quantile outputs per action, following the QR-DQN framework. This distributional representation captures return uncertainty and tail behavior, enabling the computation of CVaR-based risk-sensitive Q-values to guide policy optimization.

All actor, scalar critic, and distributional critic networks are trained using the Adam optimizer. The learning rates are set to $\omega_a = 2 \times 10^{-4}$ for the actor network and $\omega_c = 2 \times 10^{-3}$ for both scalar and distributional critics. Target networks share the same architectures as their corresponding online networks and are updated via soft updates with coefficient $\zeta = 0.005$. A complete list of training parameters is provided in Table~\ref{tab:parameters}.

For performance comparison, we consider the following representative baseline methods:
\begin{itemize}
\item \textbf{Proximal Policy Optimization (PPO)} \cite{ppo}: A  widely adopted policy-gradient-based DRL algorithm that directly learns a stochastic policy for action selection. PPO employs a clipped surrogate objective to constrain policy updates, thereby improving training stability and sample efficiency.
\item \textbf{Deep Q Network (DQN)}: A canonical value-based DRL algorithm that approximates the action-value function using deep neural networks. At each decision epoch, DQN selects actions by maximizing the estimated Q-values corresponding to the current system state.
\item \textbf{Rainbow} \cite{rainbow}: An advanced extension of DQN that combines multiple algorithmic improvements, including double Q-learning, prioritized experience replay, dueling network architectures, and multi-step bootstrapping, to improve learning stability and performance. Notably, Rainbow incorporates the C51 distributional RL component, which models the return distribution as a categorical distribution over a fixed set of discrete support atoms, estimating only the probability mass for each atom.
\item \textbf{Soft Actor-Critic (SAC)}: The baseline SAC algorithm without the diffusion-based actor or the distributional critic. 
\end{itemize}

Unless otherwise stated, all evaluations of the system average VAoI and CVaR of VAoI are conducted under a unified system configuration to ensure fair and reproducible comparisons. The system consists of $N=20$ users. For each user, packets are generated according to a Bernoulli process with rate $r_n=0.75$, and each transmission succeeds independently with probability $p=0.9$. When the scheduler transmits a packet, the transmission cost is $c_t=1$. The system is subject to a long-term average transmission cost constraint $\eta_{\max}=0.85$. During testing, the learned policy is fixed and deployed in the environment for $5000$ consecutive time slots. The resulting VAoI trajectory is recorded and used to compute the empirical average VAoI and the corresponding CVaR metrics. All baseline algorithms are evaluated under identical environmental conditions, system parameters, and testing protocols.

\subsection{Simulation Results}

\begin{figure}[t]
    \centering
    \includegraphics[width=0.95\linewidth]{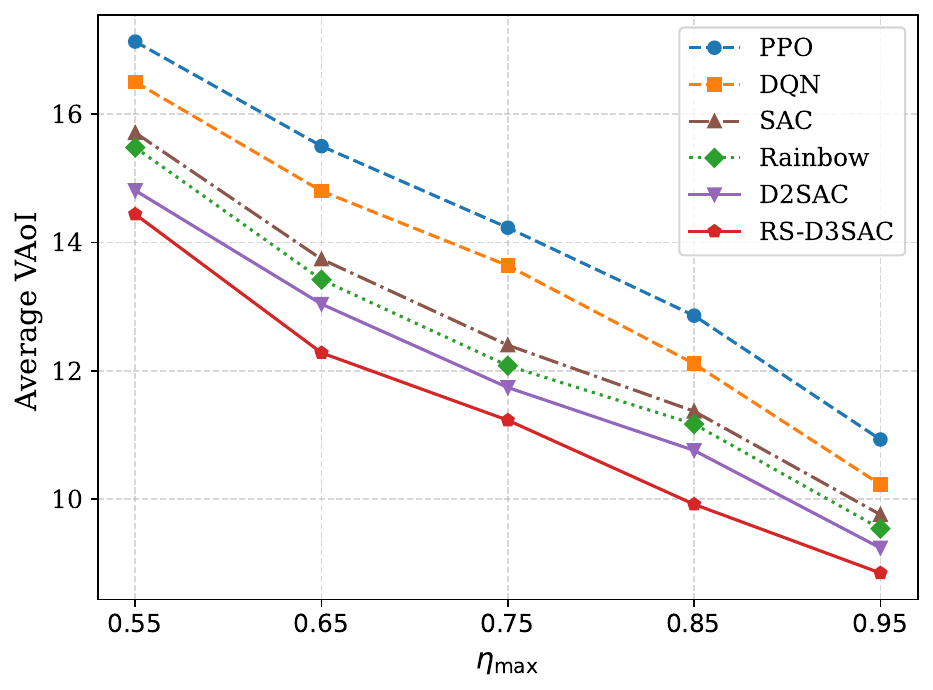}
    \caption{Comparison of system average VAoI under different transmission cost constraints: D2SAC and RS-D3SAC versus PPO, DQN, Rainbow, and SAC.}
    \label{fig:benchmark_vaoi}
    \vspace{-0.1in}
\end{figure}

\subsubsection{Benchmark Evaluation of D2SAC and RS-D3SAC}
We first evaluate the effectiveness of the proposed algorithms for VAoI scheduling by comparing the average VAoI performance of D2SAC and RS-D3SAC against the four representative DRL baselines (PPO, DQN, SAC, and Rainbow) under identical training and testing conditions. Fig.~\ref{fig:benchmark_vaoi} reports the resulting average VAoI under different long-term transmission cost constraints $\eta_{\max}$. As $\eta_{\max}$ increases, i.e., as the transmission constraint is relaxed, the average VAoI decreases for all algorithms due to the increased availability of transmission opportunities.

As shown in Fig.~\ref{fig:benchmark_vaoi}, D2SAC consistently achieves the lowest average VAoI among all baseline methods across the entire range of cost constraints. This demonstrates the effectiveness of the diffusion-based policy generation mechanism in reducing long-term information staleness. Unlike conventional policy networks that directly output actions or action probabilities, the diffusion actor generates actions through a multi-step reverse denoising process, progressively refining samples from Gaussian noise toward the optimal action distribution. This generative policy formulation enables the actor to capture complex temporal dependencies and latent system dynamics inherent in VAoI evolution, resulting in more accurate and stable scheduling decisions. As a result, D2SAC exhibits superior average VAoI performance compared with standard SAC and other conventional DRL approaches.

Although RS-D3SAC is primarily designed to mitigate tail risk rather than explicitly optimize the mean performance, its average VAoI is also reported in Fig.~\ref{fig:benchmark_vaoi} for completeness. As observed, RS-D3SAC achieves a slightly lower average VAoI than D2SAC. While RS-D3SAC incorporates risk-sensitive objectives that emphasize tail control rather than further minimizing the expected VAoI, this result indicates that introducing risk sensitivity does not come at the expense of mean performance. Instead, suppressing extreme VAoI realizations can also yield modest improvements in average performance. A more detailed discussion of this phenomenon is provided in the subsequent ablation studies.

\begin{figure}[t]
    \centering
    \includegraphics[width=0.95\linewidth]{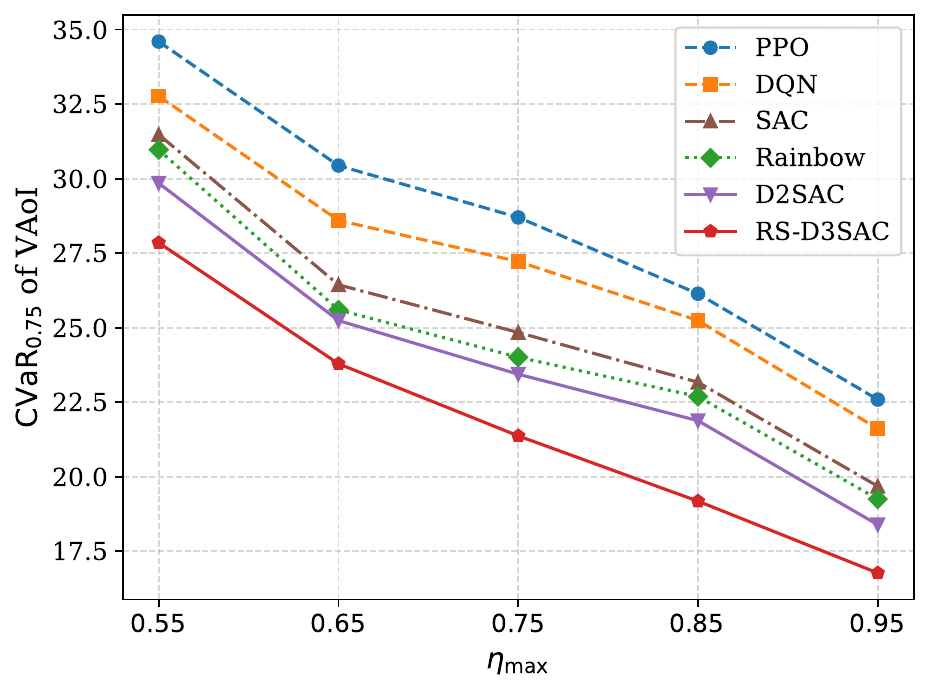}
    \caption{Comparison of system $\mathrm{CVaR}_{0.75}$ of VAoI under different transmission cost constraints: RS-D3SAC versus D2SAC, PPO, DQN, Rainbow, and SAC.}
    \label{fig:benchmark_cvar}
    \vspace{-0.1in}
\end{figure}

We next evaluate tail-risk performance using the $\mathrm{CVaR}_{0.75}$ metric, as shown in Fig.~\ref{fig:benchmark_cvar}. Similar to the average VAoI results, $\mathrm{CVaR}_{0.75}$ of VAoI decreases as $\eta_{\max}$ increases due to the relaxed resource constraint. Compared with all baseline algorithms, RS-D3SAC achieves a clear and consistent advantage in suppressing extreme VAoI values across different transmission cost constraints. 

Notably, the performance improvement of RS-D3SAC over D2SAC is more significant for CVaR than for average VAoI. This observation highlights the critical role of the distributional critic in risk-sensitive scheduling. By explicitly learning multiple quantiles of the return distribution, the distributional critic captures not only the central tendency but also the tail behavior of VAoI returns. This enables the policy optimization process to explicitly account for extreme adverse events and avoid high-risk scheduling actions, thereby achieving substantially improved tail-risk control. In contrast, D2SAC and the other baseline methods (i.e., PPO, DQN, and SAC) optimize only the expected return and therefore focus on the mean of the return distribution. As a result, they lack explicit risk-sensitive awareness of tail behavior and are inherently limited in their ability to suppress extreme VAoI outcomes under stochastic packet arrivals and transmission uncertainty.

It is also worth noting that Rainbow incorporates a distributional RL component via the C51 algorithm \cite{c51}. However, its CVaR performance remains inferior to that of RS-D3SAC. This is because C51 represents the return distribution using a fixed set of discrete atoms whose support range must be pre-specified, limiting its flexibility in accurately modeling heavy-tailed or skewed return distributions. In addition, Rainbow lacks the expressive stochastic policy representations enabled by diffusion-based actors. These limitations prevent Rainbow from fully exploiting its distributional representation for effective tail-risk mitigation. Overall, the joint integration of a diffusion-based actor and a distributional critic in RS-D3SAC yields complementary benefits, enabling strong tail-risk suppression without sacrificing average system performance.

\begin{figure}[t]
    \centering
    \includegraphics[width=0.95\linewidth]{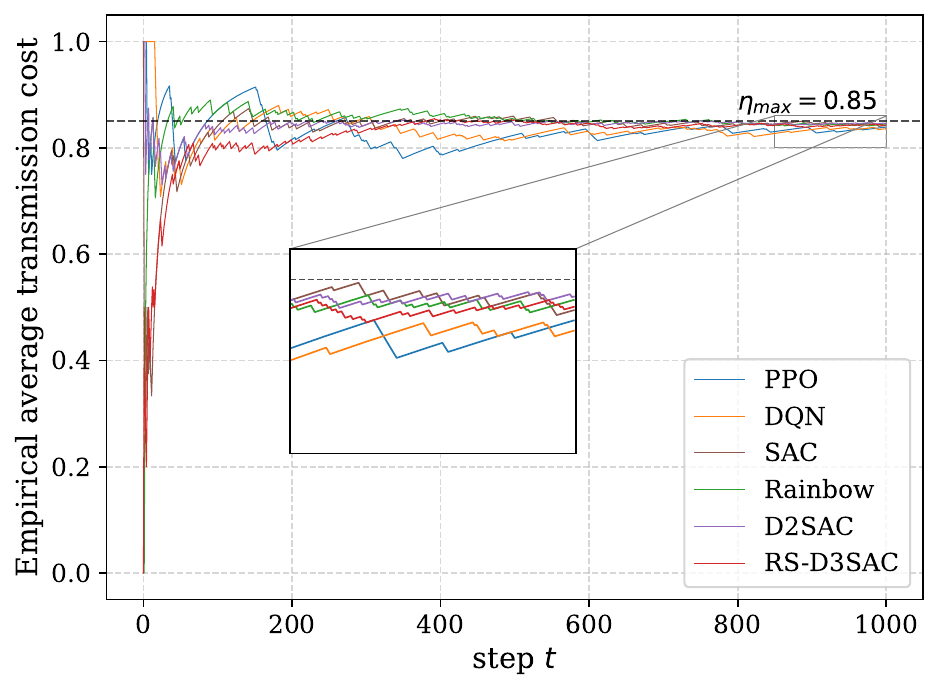}
    \caption{Empirical average transmission cost $\eta$ versus testing step $t$ under different DRL algorithms, when the performance metric is $\mathrm{CVaR}_{0.75}$ of VAoI.}
    \label{fig:cost_verification}
    \vspace{-0.1in}
\end{figure}

Fig. \ref{fig:cost_verification} illustrates the evolution of the empirical average transmission cost $\eta$ during the testing phase for all evaluated algorithms when $\mathrm{CVaR}_{0.75}$ of VAoI is considered, where the black dashed line denotes the prescribed upper bound $\eta_{\max}=0.85$. As shown, the long-term average transmission cost of each algorithm remains strictly below the constraint, confirming empirical satisfaction of the cost requirement. Recall that in the CMDP formulation (see Section \ref{sec:risk_sensitive}), the augmented reward incorporates a cost penalty weighted by the Lagrange multiplier $\lambda$, which is adaptively updated during training to balance VAoI minimization and resource compliance. This guides the learned policy toward a feasible CMDP solution. More importantly, RS-D3SAC achieves the substantial tail-risk reduction reported in Fig. \ref{fig:benchmark_cvar} without violating the transmission cost constraint. This indicates that the improved $\mathrm{CVaR}_{0.75}$ performance does not stem from increased resource consumption, but rather from a more refined optimization of the return distribution and the risk-aware Q-value evaluation within the same feasible policy set.

\begin{figure*}[htbp]
    \centering
    \begin{subfigure}[t]{0.325\textwidth}
        \includegraphics[width=\linewidth]{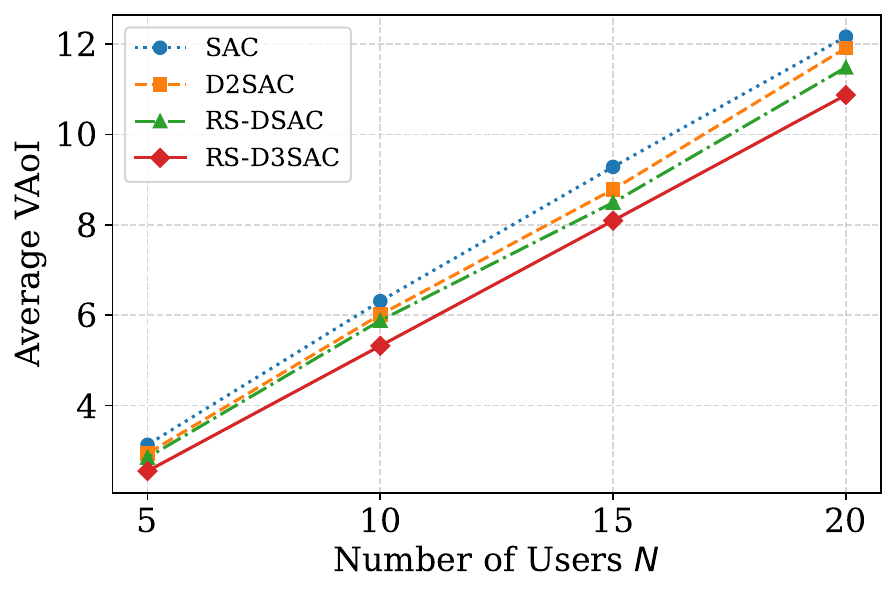}
        \caption{}
    \end{subfigure}
    \hfill
    \begin{subfigure}[t]{0.325\textwidth}
        \includegraphics[width=\linewidth]{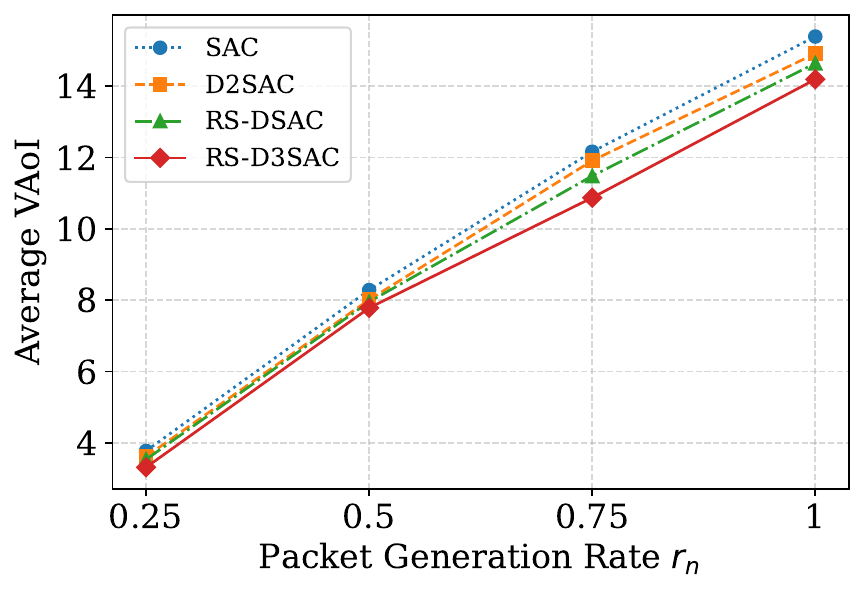}
         \caption{}
    \end{subfigure}
    \hfill
    \begin{subfigure}[t]{0.325\textwidth}
        \includegraphics[width=\linewidth]{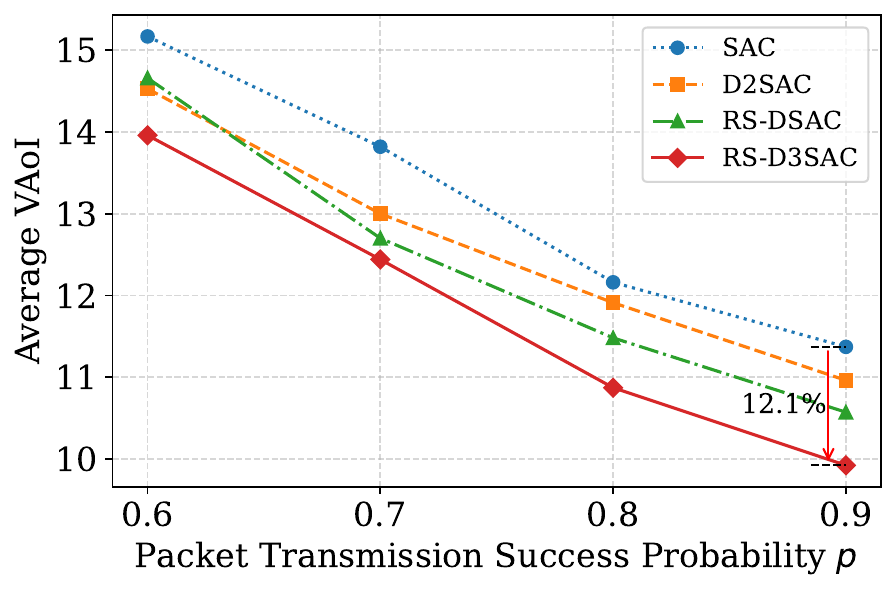}
        \caption{}
    \end{subfigure}
    \caption{Ablation study of the system-wide average VAoI performance under different system conditions. The four algorithmic variants (SAC, D2SAC, RS-DSAC, and RS-D3SAC) are compared to isolate the effects of the diffusion-based actor and the distributional critic. Subfigures show the impact of (a) number of users, (b) packet generation rate, and (c) packet transmission success probability on average VAoI.}
    \label{fig:ablation_vaoi}
     \vspace{-0.05in}
\end{figure*}

\begin{figure*}[htbp]
    \captionsetup[subfigure]{skip=1pt} 
    \centering
    \begin{subfigure}[t]{0.325\textwidth}
        \includegraphics[width=\linewidth]{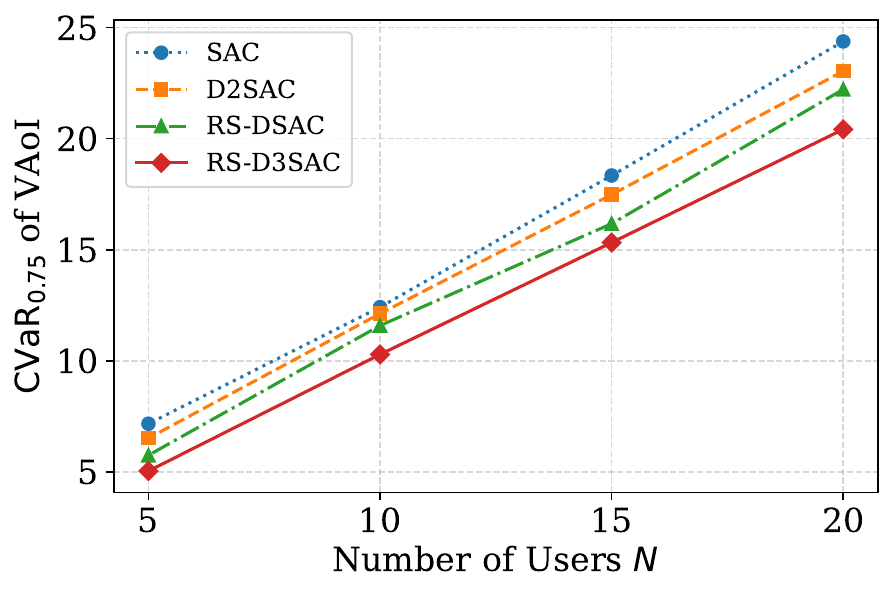}
        \caption{}
    \end{subfigure}
    \hfill
    \begin{subfigure}[t]{0.325\textwidth}
        \includegraphics[width=\linewidth]{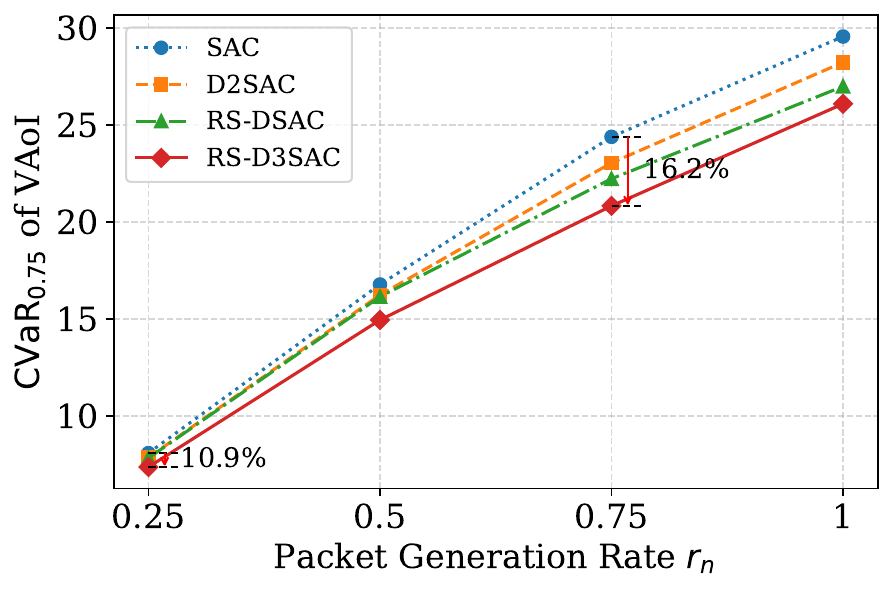}
        \caption{}
    \end{subfigure}
    \hfill
    \begin{subfigure}[t]{0.325\textwidth}
        \includegraphics[width=\linewidth]{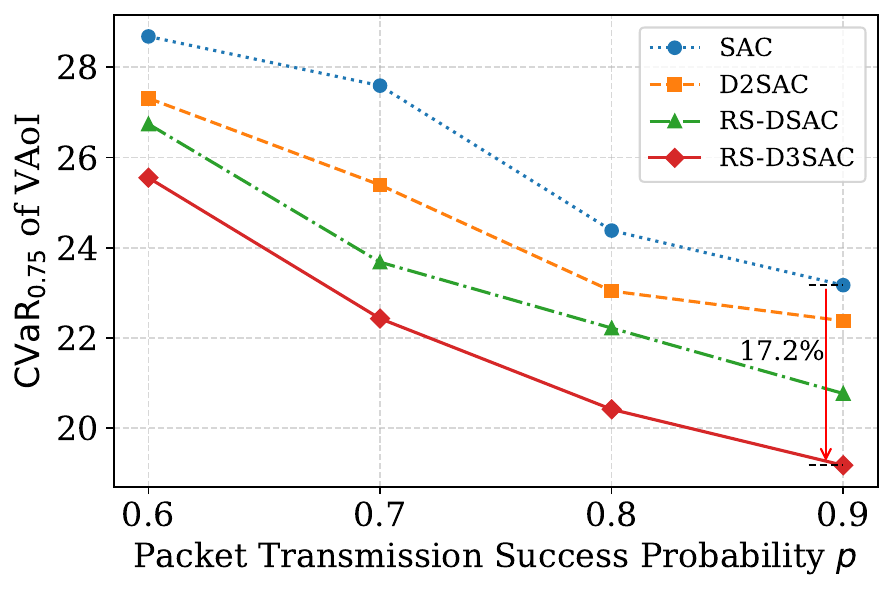}
        \caption{}
    \end{subfigure}
    \caption{Ablation study of the system-wide $\mathrm{CVaR}_{0.75}$ of VAoI performance under different system conditions. The results compare SAC, D2SAC, RS-DSAC, and RS-D3SAC to highlight the role of diffusion-based policy generation and distributional value modeling in tail-risk control. Subfigures illustrate the effect of (a) number of users, (b) packet generation rate, and (c) packet transmission success probability on tail-risk performance.}
    \label{fig:ablation_cvar}
     \vspace{-0.1in}
\end{figure*}

\subsubsection{Component-wise Ablation Analysis}
To isolate and quantify the individual contributions of the diffusion-based actor and the distributional critic, we consider four algorithmic variants: (i) SAC, serving as the baseline actor-critic method; (ii) D2SAC, which has a diffusion-based actor compared to SAC; (iii) RS-DSAC, a risk-sensitive SAC equipped with a QR-DQN-based distributional critic to compute risk-sensitive Q-value, but without the diffusion-based actor; and (iv) RS-D3SAC, which integrates both the diffusion-based actor and the risk-sensitive distributional critic.

We evaluate these algorithms along three system dimensions: (a) number of users, (b) packet generation rate, and (c) packet transmission success probability. For each setting, we report both the average VAoI and the tail-risk metric $\mathrm{CVaR}_{0.75}$, enabling a detailed comparison between mean performance and risk-sensitive behavior.

Fig. \ref{fig:ablation_vaoi} presents the average VAoI performance of the four algorithms. A consistent trend across all subfigures is that both D2SAC and RS-DSAC outperform the baseline SAC, indicating that either diffusion-based policy modeling or distributional value estimation alone can improve average performance. From a mean-performance perspective, the improvements brought by the diffusion-based actor and the distributional critic are broadly comparable, with RS-DSAC exhibiting slightly larger gains in some regimes. When tail behavior is not explicitly emphasized, both components primarily contribute by stabilizing learning and improving policy evaluation. Consequently, although RS-D3SAC achieves the lowest average VAoI overall, its incremental gain over D2SAC or RS-DSAC remains relatively modest. This observation confirms that average VAoI alone is insufficient to fully reveal the distinctive advantage of distributional value modeling.

Different insights emerge when considering tail-risk performance. Fig. \ref{fig:ablation_cvar} reports the corresponding $\mathrm{CVaR}_{0.75}$ results. In most configurations, RS-DSAC achieves substantially larger improvements over SAC than D2SAC does, clearly demonstrating the importance of explicitly modeling the return distribution for risk-sensitive optimization. Since $\mathrm{CVaR}_{0.75}$ targets the worst 25\% of outcomes, it cannot be optimized through expected returns alone. Despite its expressive diffusion-based actor, D2SAC remains mean-oriented and insensitive to tail behavior. In contrast, the distributional critic in RS-DSAC directly models multiple quantiles of the return distribution, enabling the policy to identify and suppress extreme VAoI events. Fig. \ref{fig:ablation_cvar} thus provides direct evidence that CVaR optimization intrinsically requires value distribution modeling. RS-D3SAC further combines the strengths of both components and consistently achieves the best CVaR performance; however, the dominant gain in tail-risk reduction clearly stems from the distributional critic, with the diffusion-based actor providing complementary refinement.

Figs.~\ref{fig:ablation_vaoi}(a) and \ref{fig:ablation_cvar}(a) examine performance under varying numbers of users. As the user population increases, system load intensifies, resulting in more complex scheduling decisions and heavier-tailed VAoI distributions. Consequently, the relative advantage of RS-DSAC and RS-D3SAC over SAC and D2SAC becomes more significant, particularly in terms of CVaR. This trend indicates that value distribution modeling becomes increasingly important as system scale grows. Figs.~\ref{fig:ablation_vaoi}(b) and \ref{fig:ablation_cvar}(b) illustrate the impact of packet generation rate. At low generation rates, the system operates in a lightly loaded regime, and performance differences among algorithms are relatively small. As the generation rate increases and the return distribution becomes more heavy-tailed, the benefit of distributional modeling becomes more evident. Quantitatively, the CVaR reduction achieved by RS-D3SAC compared with SAC increases from 10.9\% to 16.2\% as $r_n$ increases from $0.25$ to $0.75$, further highlighting the effectiveness of distributional critics under high-load conditions where extreme VAoI events occur more frequently. Moreover, Figs.~\ref{fig:ablation_vaoi}(c) and \ref{fig:ablation_cvar}(c) examine the impact of packet transmission success probability. RS-D3SAC consistently outperforms the other methods. For instance, when $p=0.9$, RS-D3SAC achieves a 17.2\% reduction in $\mathrm{CVaR}_{0.75}$ compared with SAC, alongside a 12.1\% reduction in average VAoI, demonstrating robust performance under stochastic channel conditions.

\begin{figure}
    \centering
    \includegraphics[width=0.95\linewidth]{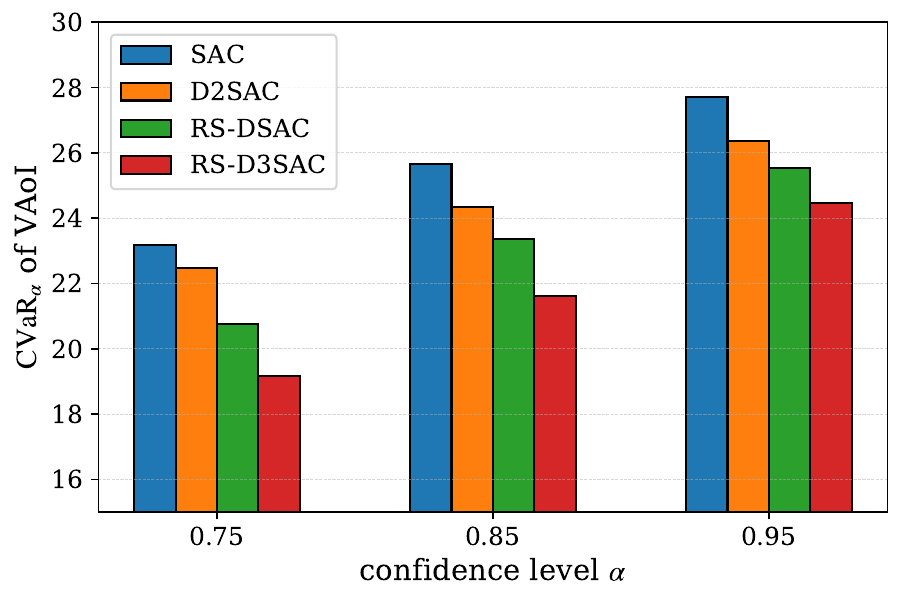}
    \caption{Performance comparison of SAC, D2SAC, RS-DSAC, and RS-D3SAC under different confidence levels.}
    \label{fig:cvar_risk}
    \vspace{-0.1in}
\end{figure}

Fig.~\ref{fig:cvar_risk} illustrates the tail-risk performance of the SAC-based algorithms for varying CVaR confidence levels. As $\alpha$ increases, the CVaR of VAoI rises accordingly, since larger confidence levels place greater emphasis on rare but severe VAoI realizations. Across all confidence levels, RS-DSAC consistently outperforms D2SAC, demonstrating the advantage of explicitly learning the return distribution rather than relying on expectation-based value estimates. Thanks to the synergistic integration of the diffusion-based actor and the distributional critic, RS-D3SAC remains effective across a broad spectrum of risk sensitivities, maintaining superior performance even under stringent tail-risk requirements, such as $\alpha=0.95$. These results further confirm that RS-D3SAC provides a robust and scalable solution for risk-sensitive VAoI scheduling, demonstrating its effectiveness in controlling extreme semantic staleness events under stringent risk requirements.

\section{Conclusion}\label{sec:conclusion}
We have investigated both average-oriented and tail-risk-sensitive VAoI scheduling in multi-user wireless status update systems with stochastic packet arrivals, unreliable channels, and long-term transmission cost constraints. In particular, we put forth RS-D3SAC, a risk-sensitive deep distributional diffusion-based SAC algorithm that integrates a diffusion-based actor with a distributional critic to enable effective tail-risk-aware control.

To establish a baseline, we first formulated average VAoI minimization as a CMDP and solved it using the deep diffusion-based SAC (D2SAC) algorithm, which leverages a diffusion process to enhance policy expressiveness and achieve superior mean performance. Building on this foundation, RS-D3SAC replaces the scalar critic with a quantile-based distributional critic that explicitly models the full VAoI return distribution. Tail-risk-sensitive Q-values derived from the CVaR metric then guide policy improvement, allowing the actor to prioritize worst-case VAoI reduction while meeting long-term transmission cost constraints.

Extensive simulations demonstrate that D2SAC substantially improves average VAoI over standard SAC and other DRL benchmarks. More importantly, RS-D3SAC achieves marked reductions in system CVaR of VAoI. The dominant gain in tail-risk control originates from the distributional critic, with the diffusion-based actor providing complementary gains to stabilize and enrich policy learning. These results highlight the effectiveness of combining diffusion-based policy learning with distributional value estimation, establishing RS-D3SAC as a robust and principled framework for risk-aware VAoI scheduling in complex wireless networks.

\ifCLASSOPTIONcaptionsoff
  \newpage
\fi



\bibliographystyle{IEEEtran}
\bibliography{IEEEabrv,main}
%

%








\end{document}